\documentclass[lettersize,journal]{IEEEtran}
\usepackage{amsmath}
\usepackage{amssymb}

\usepackage{graphicx}
\usepackage{epstopdf}
\usepackage{array}
\newcommand{\PreserveBackslash}[1]{\let\temp=\\#1\let\\=\temp}
\newcolumntype{C}[1]{>{\PreserveBackslash\centering}p{#1}}
\newcolumntype{R}[1]{>{\PreserveBackslash\raggedleft}p{#1}}
\newcolumntype{L}[1]{>{\PreserveBackslash\raggedright}p{#1}}
\usepackage[usenames]{color}
\usepackage{colortbl,booktabs}
\usepackage{tabularx}
\usepackage{multicol}
\usepackage{booktabs}
\usepackage[Symbol]{upgreek}
\usepackage{bm}
\usepackage{cite}
\usepackage{balance}
\usepackage{mathrsfs}
\usepackage{url}
\usepackage{threeparttable}
\usepackage{dcolumn}
\usepackage{multirow}
\usepackage{booktabs}
\usepackage{amsfonts}
\usepackage{algorithm}
\usepackage{algorithmic} 
\usepackage{float}
\usepackage{subfigure} 
\usepackage{cuted} 
\begin{document}
	\title{Training Beam Design for Channel Estimation in Hybrid mmWave MIMO Systems}
	\author{Xiaochun Ge,~\IEEEmembership{}
			Wenqian Shen,~\IEEEmembership{}
			Chengwen Xing,~\IEEEmembership{Member, IEEE, }		
			Lian Zhao,~\IEEEmembership{Senior Member, IEEE, }
			and Jianping An,~\IEEEmembership{Member, IEEE}
			\thanks{
			This work was supported in part by the National Natural Science Foundation of China (NSFC) under Grant 61901034, and in part by NSFC under Grant U1836201, NSFC under Grant 62071398, and NSFC under Grant 62101614. 
			This work was also supported in part by Open Research Fund of the Shaanxi Province Key Laboratory of Information Communication Network and Security under Grant ICNS201905 and Ericsson. \textit{(Corresponding author: Wenqian Shen.)}
			
			X. Ge, W. Shen, C. Xing, and J. An are with the School of Information and Electronics, Beijing Institute of Technology, Beijing 100081, China (e-mails: xiaochun\_ge\_bit\_ee@163.com; shenwq@bit.edu.cn; xingchengwen@gmail.com; an@bit.edu.cn). 
			
			L. Zhao is with the Department of Electrical, Computer, and Biomedical Engineering, Ryerson University, Toronto, ON M5B 2K3, Canada (e-mail: l5zhao@ryerson.ca).}
			}

	\maketitle
	\begin{abstract}
		Training beam design for channel estimation with infinite-resolution and low-resolution phase shifters (PSs) in hybrid analog-digital milimeter wave (mmWave) massive multiple-input multiple-output (MIMO) systems is considered in this paper.
		By exploiting the sparsity of mmWave channels, the optimization of the sensing matrices (corresponding to training beams) is formulated according to the compressive sensing (CS) theory.
		Under the condition of infinite-resolution PSs, we propose relevant algorithms to construct the sensing matrix, where the theory of convex optimization and the gradient descent in Riemannian manifold is used to design the digital and analog part, respectively.
		Furthermore, a block-wise alternating hybrid analog-digital algorithm is proposed to tackle the design of training beams with low-resolution PSs, where the performance degeneration caused by non-convex constant modulus and discrete phase constraints is effectively compensated to some extent thanks to the iterations among blocks.		
		Finally, the orthogonal matching pursuit (OMP) based estimator is adopted for achieving an effective recovery of the sparse mmWave channel. 		
		Simulation results demonstrate the performance advantages of proposed algorithms compared with some existing schemes.
	\end{abstract}
	
	\begin{IEEEkeywords}
		Hybrid mmWave MIMO, channel estimation, training beam design, sensing matrix.
	\end{IEEEkeywords}
	\IEEEpeerreviewmaketitle
	
	\section{Introduction}\label{S1}
	\IEEEPARstart{A}{s} 
	one of the key techniques in the future mobile communication systems \cite{Background_nextgenMIMO_Larsson_IEEECOMM_2014}, milimeter wave (mmWave) communication technique has attracted extensive attentions in recent years.
	The bandwidth of mmWave can reach more than 2 GHz and it can support data transmission at the rate of 10 Gbps, making it possible to construct high-speed broadband communication systems \cite{Background_Itwillwork_Heath_IEEEaccess_2013}.
	Besides, by taking the advantage of short wavelength of mmWave, large-scale antenna arrays can be integrated in a limited space and further realize high beamforming gain to compensate for the high path loss of mmWave channels \cite{Background_Overview_Heath_JSTSP_2016,Background_Channel_TAT_2013}.

	However, there are some technical problems in mmWave massive multiple-input multiple-output (MIMO) systems. 
	For example, the large number of antennas can cause unbearable power consumption, making it difficult to implement traditional full-digital precoding \cite{Background_Hybrid_Gaoxinyu_2018,Background_largeMIMO_Larsson_IEEESPM_2013,Background_IntrommWave_IEEECOMM_2011}.
	For this purpose, a hybrid precoding structure consisting of two parts has been proposed, namely a low-dimensional digital part which requires a small number of RF chains and an analog part implemented by a phase shifter (PS) network \cite{Matrix-Monotonic_BITXing_TSP_2019,Major-Min_BITXing_TSP_2020,Training-Opt_BITXing_TWC_2020}.
	However, the high-resolution PSs can still cause great energy consumption and hardware costs \cite{Background_Overview_Heath_JSTSP_2016}.
	To effectively remedy this situation while ensuring sufficient system performance, a mmWave hybrid analog-digital structure based on low-resolution PSs has been proposed \cite{Background_RFswitch_Heath_Access_2016,LowPSPrecoding_YuWei_JSTSP_2015}.

	The main difficulty of hybrid precoding lies in the constant modulus constraints of analog precoder, while the equipment of low-resolution PSs additionally adds discrete phase constraints on analog precoder design, making the hybrid precoding problem more intractable \cite{HybridPrecoding_OmarEl_TWC_2014}.
	Therefore, the issue of hybrid analog-digital precoding with low-resolution PSs has been studied in \cite{LowPSPrecoding_Taiwan_TVT_2017,LowPSPrecoding_YuWei_JSTSP_2015, LowPSPrecoding_DaGong_TSP_2018}, 
	where the authors in \cite{LowPSPrecoding_Taiwan_TVT_2017} aimed for minimizing the distance between full-digital precoding matrix and hybrid precoding matrix in point-to-point communication systems under low-resolution PSs conditions.
	Sohrabi and Yu directly began with maximizing the spectral efficiency, and proposed an iterative algorithm under low-resolution PSs conditions to iteratively design the analog precoding matrix column-by-column \cite{LowPSPrecoding_YuWei_JSTSP_2015}.
	In \cite{LowPSPrecoding_DaGong_TSP_2018}, Wang et al. further investigated the design of hybrid precoder in an extreme case where the resolution of PSs was reduced to one bit.

	However, all the aforementioned studies are based on an assumption that the channel state information (CSI) is known. 
	When the CSI is unknown or inaccurate, the system performance will be deteriorated seriously.
	Unfortunately, the large number of antennas in massive MIMO systems and the low signal-to-noise ratio (SNR) before beamforming bring a great challenge to channel estimation \cite{CE_Ahmed_JSTSP_2014}.
	Compressive sensing (CS) theory can be used to proceed a large-scale channel estimation with small-scale measurements by leveraging the sparsity of mmWave channels.
	The design of sensing matrices is of great importance, which is closely related to the performance of sparse signal recovery \cite{CSBook_Intro}.

	Literatures \cite{SensingOpti_CSAlgorithmAnalysis_TIT_2004, SensingOpti_OMPAnalysis_TIT_2011, SensingOpti_JointOptimization_TIP_2009, 
	SensingOpti_AlterOptimization_ZheGong_TSP_2015, SensingOpti_SVD_ZheGong_TSP_2013,SensingOpti_Block_Israel_TSP_2011} investigated some relevant techniques about the sensing matrix design.		
	In \cite{SensingOpti_CSAlgorithmAnalysis_TIT_2004, SensingOpti_OMPAnalysis_TIT_2011}, 
	by jointly analyzing several OMP-based sparse signal recovery algorithms and sensing matrix design algorithms, the authors summarized different parameters related to the coherence level of equivalent dictionary matrices, where the parameters investigated were of great significance to the training beam design.
	The authors in \cite{SensingOpti_JointOptimization_TIP_2009,
	SensingOpti_AlterOptimization_ZheGong_TSP_2015} considered the joint/alternating optimization of sensing matrix and dictionary matrix.
	In \cite{SensingOpti_SVD_ZheGong_TSP_2013}, the sensing matrix was optimized with a fixed dictionary matrix, and the optimal closed-form solution of sensing matrix was obtained.
	Furthermore, a low-complexity implementation based on eigenvalue decomposition for the sensing matrix was given in \cite{SensingOpti_Block_Israel_TSP_2011}.
	However, none of literatures \cite{SensingOpti_JointOptimization_TIP_2009, 
	SensingOpti_AlterOptimization_ZheGong_TSP_2015, SensingOpti_SVD_ZheGong_TSP_2013,SensingOpti_Block_Israel_TSP_2011} is suitable for the case where the designed sensing matrix is composed of analog and digital parts (corresponding to the hybrid beam training in mmWave MIMO systems).
	The authors in \cite{ManifoldOpti_AltMin_JSTSP_2016} first attempted to directly solve the hybrid precoder design problem by using the Riemannian manifold, and more relevant techniques can be found in 
	\cite{ManiOptBook_1_Optimization_Algorithms_on_Matrix_Manifolds, ManiOptBook_2_Manifold_Learning_Theory_and_Applications, ManiOpt_applyinComm_Gangke_ISIT_2015, ManiOpt_Toolbox_ML_2014}.
	Though the application scenarios and optimization goals are different, literatures \cite{ManifoldOpti_AltMin_JSTSP_2016,ManiOptBook_1_Optimization_Algorithms_on_Matrix_Manifolds, ManiOptBook_2_Manifold_Learning_Theory_and_Applications, ManiOpt_applyinComm_Gangke_ISIT_2015, ManiOpt_Toolbox_ML_2014} inspire us to make full use of the powerful tool of manifold optimization. 
	Therefore, the theory of manifold optimization is adopted in this paper to tackle the constant modulus constraints imposed on analog sensing matrices. 

	There is very little existing research focusing on the hybrid analog-digital sensing matrix design for CS-based channel estimation problems in massive MIMO systems.
	A CS-based sparse signal recovery problem for channel estimation was formulated in \cite{CE_Model_Korea_GCC_2014,CE_OMP_Korea_TCOM_2016}. 
	However, the authors only consider the case where the analog sensing matrix is square (i.e. the number of training beams is equal to the number of antennas), and set it as a DFT matrix.
	The more general case in massive MIMO where the number of training beams is less than that of antennas is not considered.
	The authors in \cite{LowPSCE_BeiDa_VTC_2017} focus on the channel estimation problems in wideband mmWave systems.
	Although the application scenarios and system models are different from those in this paper, the random scheme (the low-resolution analog sensing matrix is generated randomly and uniformly with quantized phases) given in \cite{LowPSCE_BeiDa_VTC_2017} also provides an idea for the analog sensing matrix design.
	The randomness causes inevitable performance loss, so it is necessary to propose an effective sensing matrix design in infinite-resolution and low-resolution PSs based mmWave massive MIMO systems with a hybrid analog-digital structure.

	Our contributions can be summarized as follows:
	\begin{itemize}
		\item[$\bullet$] We transform the problem of training beam design for channel estimation into the problem of hybrid sensing matrix design with a given dictionary matrix based on the theory of compressive sensing.
		A more generalized model is considered based on some existing papers, where the number of training beams and data streams is flexible, and the adjusted problem formulation is provided, accordingly.
		\item[$\bullet$] An alternating hybrid analog-digital sensing matrix design algorithm with infinite-resolution PSs is proposed, where the Riemannian manifold is introduced for dealing with the non-convex constant modulus constraints imposed on the analog sensing matrix.
		Besides, a joint optimization of multiple blocks is proposed for the block-diagonal digital sensing matrix design, where the original non-convex problem is transformed into a convex one by introducing an intermediate variable, and the digital sensing matrix is obtained using the eigenvalue decomposition.
		\item[$\bullet$] For the low-resolution PSs scenario, a block-wise optimization form is formulated to effectively compensate for the performance loss caused by discrete phase determined by the low-resolution PSs. 
		An alternating algorithm is proposed accordingly, where the analog sensing submatrix is obtained in the Riemannian manifold with a subsequent phase quantization step, and the digital sensing submatrix is obtained by using the gradient descent algorithm with an adaptive stepsize.
		\item[$\bullet$] Finally, we provide the numerical results and performance analysis of the proposed methods for both infinite-resolution and low-resolution PSs scenarios.
		Aside of the comparison with some existing schemes, the convergence of proposed algorithms is also verified by the simulation results.
		Furthermore, we compare the spectral efficiency of hybrid precoding using the channel estimated by different schemes, where the hybrid precoder and combiner are designed according to existing literatures.
		Numerical results show the superior performance of the algorithms we proposed.
	\end{itemize}

	The rest of the paper is organized as follows. 
	The system model is presented in Section \ref{S2}, and a further formulation for hybrid sensing matrix design is given in Section \ref{S3}.
	Then, two alternating hybrid analog-digital sensing matrix design algorithms with infinite-resolution and low-resolution PSs are proposed in Section \ref{S4} and Section \ref{S5}, respectively.
	We provide our simulation results in Section \ref{S6}.
	Finally, our conclusions are drawn in Section \ref{S7}.

	\emph{Notation}: 
	Lower-case and boldface capital letters represent column vectors and matrices, respectively.
	${\rm{vec}}\left(  \cdot  \right)$ denotes the vectorization of a matrix and ${\rm{invec}}\left(  \cdot  \right)$ denotes the invectorization of a vector.
	We let $(\cdot)^{*}$, $(\cdot)^{T}$, $(\cdot)^{H}$, $(\cdot)^{-1}$, ${\left\|  \cdot  \right\|_p}$, $\mathbb{E} \left\{  \cdot  \right\}$, $\Re \left\{  \cdot  \right\}$ and ${\rm{Tr}}\left\{  \cdot  \right\}$ 
	denote the conjugate, transpose, conjugate transpose, inverse, $p$-norm, statistical expectation, real
	part and trace operators, respectively.
	$\left|  \cdot  \right|$ denotes the determinant of a matrix or the absolute value of a scalar.
	The operator $\otimes$ and $\circ $ represent the Kronecker and Hadamard products, respectively.
	${\bf{I}}_M$ denotes an identity matrix of size $M \times M$, while ${\bf{1}}_{M,N}$ and ${\bf{0}}_{M,N}$ denote an all 1 matrix and an all 0 matrix of size $M \times N$, respectively.
	${\cal C}{\cal N}({\bf{0}},{\bf{R}})$ represents the zero-mean complex Gaussian distribution with covariance matrix $\bf{R}$.
	
	\section{System Model}\label{S2}

	\begin{figure*}[t]
		\center{\includegraphics[width=1.995\columnwidth]{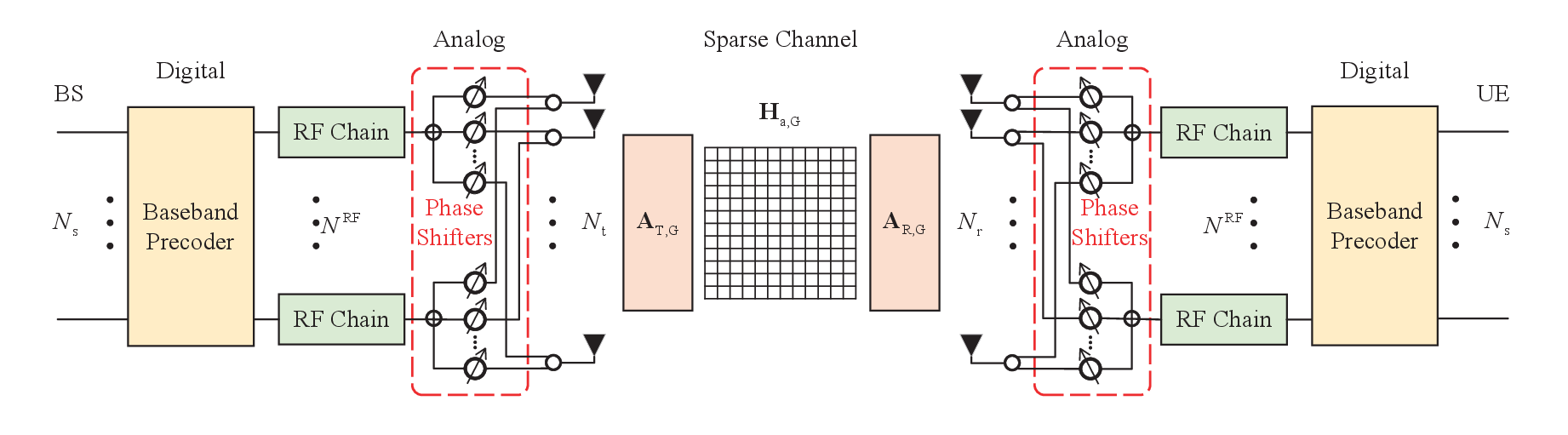}}
		\caption{Point-to-point mmWave massive MIMO system with a hybrid analog-digital structure.}
		\label{fig1}
	\end{figure*}

	In this section, for a point-to-point mmWave MIMO system, we present the system model and channel model for downlink channel estimation. 

	\subsection{Point-to-point MmWave MIMO System Model}\label{S2.1}

	As shown in Fig. \ref{fig1}, we consider a point-to-point mmWave MIMO system with a hybrid analog-digital structure, where the analog part is implemented by PSs.
	There are $N_\text{t}$ and $N_\text{r}$ antennas equipped at the base station (BS) and the user equipment (UE), respectively, and there are $N^\text{RF}$ radio frequency (RF) chains at both sides, where $N^\text{RF} \ll N_\text{t}$, $N^\text{RF} \ll N_\text{r}$.
	We assume that the number of transmitting data streams is less than or equal to that of RF chains, i.e. ${N}_{\text{s}} \leq  N^{\text{RF}}$.
	In this paper, for achieving an accurate channel estimation, ${T_\text{t}}$ training beams need to be transmitted at BS and receiving signals need to be combined using ${T_\text{r}}$ training beam patterns at UE.
	Since the number of transmitting and receiving training beams within a time slot is limited, here we divide ${T_\text{t}}$ transmitting training beams into $N_\text{t}^\text{Block}$ blocks, where each block contains $N_\text{s}$ beams to be transmitted within one time slot, i.e. ${T_\text{t}=N_{\text{s}}N_{\text{t}}^{\text{Block}}}$. 
	At the receiver, ${T_\text{r}}$ combining beams are divided into $N_\text{r}^\text{Block}$ blocks, where each block contains $N_\text{s}$ combining beams to be used in one time slot, i.e. ${T_\text{r}=N_{\text{s}}N_{\text{r}}^{\text{Block}}}$.

	Taking the $p$-th $\left( p=1,2,...,N_{\text{t}}^{\text{Block}} \right)$ block at BS and the $q$-th $\left( q=1,2,...,N_{\text{r}}^{\text{Block}} \right)$ block at UE as an example, the process of beam training can be described as follows.
	First, the transmitting pilot vector ${\bf{s}\in {{\mathbb{C}}^{N_{\text{s}}\times 1}}}$ is precoded by a baseband precoder ${{\mathbf{F}}_{\text{BB},p}\in {{\mathbb{C}}^{N^{\text{RF}}\times N_{\text{s}}}}}$.
	Then, the digital precoded signal is mapped to transmitting antennas by an analog precoder ${{\bf{F}}_{{\rm{RF}},p}} \in {{\mathbb{C}}^{{N_{\rm{t}}} \times {N^{{\rm{RF}}}}}}$.
	The procedure at UE is similar, we first combine the signal received by antennas using an analog combiner ${{\bf{W}}_{{\rm{RF}},q}} \in {{\mathbb{C}}^{{N_{\rm{r}}} \times {N^{{\rm{RF}}}}}}$.
	Then, the combined signal is further processed by a baseband (digital) combiner ${{\mathbf{W}}_{\text{BB},q} \in {{\mathbb{C}}^{N^{\text{RF}}\times N_{\text{s}}}}}$.
	By concatenating $N_{\text{s}}$ successive transmitting beam patterns in the $p$-th transmitting block and corresponding combining vectors, the resultant received signal ${{\mathbf{Y}}_{q,p}}$ of dimension $N_{\text{s}}\times N_{\text{s}}$ can be written as
	\begin{align}\label{single_block}
    {{\bf{Y}}_{q,p}} \!=\!\! {\bf{W}}_{{\text{BB}},q}^H\!{\bf{W}}_{{\text{RF}},q}^H{\bf{H}}{{\bf{F}}_{{\text{RF}},p}}{{\bf{F}}_{{\text{BB}},p}}{{\bf{S}}_p} \!\!+\!\! {\bf{W}}_{{\text{BB}},q}^H{\bf{W}}_{{\text{RF}},q}^H{{\bf{N}}_{q,p}},\!
	\end{align}
	where $\mathbf{H}\in {{\mathbb{C}}^{{{N}_{\text{r}}}\times {{N}_{\text{t}}}}}$ denotes the channel matrix, the columns of noise matrix ${{\bf{N}}_{q,p}} \in {{\mathbb{C}}^{{N_{\text{r}}} \times N_{{\text{s}}}}}$ are Gaussian noise vectors with $\mathcal{C}\mathcal{N}\left( 0,\sigma _{n}^{2}{{\mathbf{I}}_{{{N}_{\text{r}}}}} \right)$, and ${{\bf{S}}_p} \in {{\mathbb{C}}^{{N_{\text{s}}} \times {N_{\text{s}}}}}$ is a superposed matrix carrying $N_{\text{s}}$ pilot vectors.

	To simplify the expression, we assume identical pilot matrices so that ${{\mathbf{S}}_{{p}}}=\sqrt{P}{{\mathbf{I}}_{N_{\text{s}}}}$ \cite{CE_OMP_Korea_TCOM_2016}, where $P$ is the power of transmitting pilots. Then, the received signal in (\ref{single_block}) can be rewritten as
	\begin{align}\label{}
    {{\bf{Y}}_{q,p}} = \sqrt P {\bf{W}}_{{\text{BB}},q}^H{\bf{W}}_{{\text{RF}},q}^H{\bf{H}}{{\bf{F}}_{{\text{RF}},p}}{{\bf{F}}_{{\text{BB}},p}} + {{{\bf{\tilde N}}}_{q,p}},
	\end{align}
	where ${{{\bf{\tilde N}}}_{q,p}} = {\bf{W}}_{{\text{BB}},q}^H{\bf{W}}_{{\text{RF}},q}^H{{\bf{N}}_{q,p}}$ denotes the equivalent noise matrix.
	Furthermore, training beams in different transmitting blocks can be concatenated, and the same applies to combining beams in different combining blocks. 

	By concatenating the $N_{\text{t}}^{\text{Block}}$ transmitting blocks and $N_{\text{r}}^{\text{Block}}$ combining blocks, the aggregated received signal $\mathbf{Y} \in {{\mathbb{C}}^{{T_\text{r}}\times {T_\text{t}}}}$ can be given by
	\begin{align}\label{hebingxingshi}
		\mathbf{Y}=\sqrt{P}\mathbf{W}_{\text{BB}}^{H}\mathbf{W}_{\text{RF}}^{H}
		\mathbf{H}{{\mathbf{F}}_{\text{RF}}}{{\mathbf{F}}_{\text{BB}}}+\widetilde{\mathbf{N}},
	\end{align}
	where we define $M_\text{t}=N_\text{RF} N_{\text{t}}^{\text{Block}}$, $M_\text{r}=N_\text{RF} N_{\text{r}}^{\text{Block}}$, then
	${{\mathbf{F}}_{\text{RF}}}=\left[ {{\mathbf{F}}_{\text{RF},1}},...,{{\mathbf{F}}_{\text{RF},N_{\text{t}}^{\text{Block}}}} \right]\in {{\mathbb{C}}^{{{N}_{\text{t}}}\times {M_\text{t}}}}$ and ${{\mathbf{W}}_{\text{RF}}}=\left[ {{\mathbf{W}}_{\text{RF},1}},...,{{\mathbf{W}}_{\text{RF},N_{\text{r}}^{\text{Block}}}} \right]\in {{\mathbb{C}}^{{{N}_{\text{r}}}\times {M_\text{r}}}}$ are the aggregated analog precoder and combiner, respectively.
	The block-diagonal digital precoder and combiner can be expressed as ${{\mathbf{F}}_{\text{BB}}}=\text{diag}({{\mathbf{F}}_{\text{BB},1}},...,{{\mathbf{F}}_{\text{BB},N_{\text{t}}^{\text{Block}}}})\in {{\mathbb{C}}^{{M_\text{t}}\times {T_\text{t}}}}$ and ${{\mathbf{W}}_{\text{BB}}}=\text{diag}({{\mathbf{W}}_{\text{BB},1}},...,{{\mathbf{W}}_{\text{BB},N_{\text{r}}^{\text{Block}}}})\in {{\mathbb{C}}^{{M_\text{r}}\times {T_\text{r}}}}$, respectively.
	For $p=1,2,...,N_{\text{t}}^{\text{Block}}$, we define $ {{{\bf{Y}}}_p} \!=\! {\left[ {{\bf{Y}}_{1,p}^T,\!...,\!{{{\bf{Y}}}_{N_{\text{r}}^{{\text{Block}}},p}^T}} \right]^T}\!\in\!{{\mathbb{C}}^{{T_\text{r}}\times {N_\text{s}}}}$ and $ {{{\bf{\tilde N}}}_p} \!=\! {\left[ {{\bf{\tilde N}}_{1,p}^T,...,{{{\bf{\tilde N}}}_{N_{\text{r}}^{{\text{Block}}},p}^T}} \right]^T}\in{{\mathbb{C}}^{{T_\text{r}}\times {N_\text{s}}}}$, then we have $\mathbf{Y}\!=\!\left[ {{{\mathbf{Y}}}_{1}},\!...,\!{{{\mathbf{Y}}}_{N_{\text{t}}^{\text{Block}}}} \right]$, and the resultant equivalent noise matrix can be given by $\mathbf{\tilde{N}}=\left[ {{{\mathbf{\tilde{N}}}}_{1}},...,{{{\mathbf{\tilde{N}}}}_{N_{\text{t}}^{\text{Block}}}} \right]\in {{\mathbb{C}}^{{T_\text{r}}\times {T_\text{t}}}}$.

	Besides, the elements of analog precoding and combining matrices implemented by PSs should satisfy constant modulus and discrete phase constraints.
	Therefore, we have ${{\mathbf{F}}_{\text{RF}}}(i,j)\in \mathcal{F}\triangleq \left\{{{e}^{j\frac{2\pi b}{{{2}^{B}}}}}\mid b=1,2,... ,{{2}^{B}} \right\}$ and ${{\mathbf{W}}_{\text{RF}}}(i,j)\in \mathcal{W}\triangleq \left\{{{e}^{j\frac{2\pi b}{{{2}^{B}}}}}\mid b=1,2,... ,{{2}^{B}} \right\}$, where $B$ denotes the resolution (bits) of PSs.
	
	\subsection{MmWave Channel Model}\label{S2.2}

	The high path loss of propagation in space at mmWave frequency greatly limits the spatial selectivity, making the mmWave channel sparse in spatial domain. 
	Moreover, the mmWave transmitter and receiver are often equipped with densely packed large-scale antenna arrays, making different antennas highly correlated \cite{Background_IntrommWave_IEEECOMM_2011,HybridPrecoding_OmarEl_TWC_2014,Background_Channel_TAT_2013}. 
	The Saleh-Valenzuela model was proposed based on aforementioned characteristics, where the channel matrix can be expressed as the sum of $L$ propagation paths \cite{CE_Ahmed_JSTSP_2014,MIMOBook_Dailinglong}, which can be given by
	\begin{align}\label{channel_Saleh-Valenzuela}
    {\bf{H}} = \sqrt {\frac{{{N_{\text{t}}}{N_{\text{r}}}}}{L}} \sum\limits_{l = 1}^L {{\alpha _l}} {{\bf{a}}_{\text{r}}}\left( {{\varphi _l}} \right){\bf{a}}_{\text{t}}^H\left( {{\phi _l}} \right),
	\end{align}
	where ${\alpha _l}\sim ~{\mathcal C}{\mathcal N}\left( {0,\frac{1}{L}} \right)$ represents the complex gain of the $l$-th path, $\varphi _{l}$ and $\phi _{l}$ represent the azimuth angles of arrival and departure (AoAs/AoDs), respectively.
    The steering vectors ${{\mathbf{a}}_{\text{r}}}\left( \varphi _{l} \right)$ and ${{\mathbf{a}}_{\text{t}}}\left( \phi _{l} \right)$ denote the normalized receiving/transmitting array response with the azimuth angle of $\varphi _{l}$ and $\phi _{l}$, respectively, which are closely related to the array structure at both transmitting and receiving sides.
	We consider that uniform linear arrays (ULAs) are equipped at both BS and UE, and the corresponding array response vectors can be expressed as
	\begin{align}\label{}
    {{\mathbf{a}}_{\text{t}}}\left( \phi _{l} \right)\!=\!\!\frac{1}{\sqrt{{{N}_{\text{t}}}}}{{\left[ 1,\!{{e}^{j2\pi d\cos \left( \phi _{l} \right)/\lambda }},\!...,\!{{e}^{j2\pi \left( {{N}_{\text{t}}}-1 \right)d\cos \left( \phi _{l} \right)/\lambda }} \right]}^{\!T}}\!\!,\!
	\end{align}
	\begin{align}\label{}
    {{\mathbf{a}}_{\text{r}}}\left( \varphi _{l} \right)\!=\!\!\frac{1}{\sqrt{{{N}_{\text{r}}}}}{{\left[ 1,\!{{e}^{j2\pi d\cos \left( \varphi _{l} \right)/\lambda }},\!...,\!{{e}^{j2\pi \left( {{N}_{\text{r}}}-1 \right)d\cos \left( \varphi _{l} \right)/\lambda }} \right]}^{\!T}}\!\!,\!\!
	\end{align}
	where $\lambda$ denotes the wavelength of the signal and $d=\frac{\lambda}{2}$ is the commonly used interelement spacing of antennas \cite{CE_OMP_Korea_TCOM_2016}.

    We can rewrite the channel matrix in (\ref{channel_Saleh-Valenzuela}) in a more compact form as
    \begin{align}\label{}
    \mathbf{H}={{\mathbf{A}}_{\text{R}}}{{\mathbf{H}}_{\text{a}}}\mathbf{A}_{\text{T}}^{H},
	\end{align}
	where we have ${{\mathbf{A}}_{\text{T}}}=\left[ {{\mathbf{a}}_{\text{t}}}\left( \phi _{1} \right),{{\mathbf{a}}_{\text{t}}}\left( \phi _{2} \right),... ,{{\mathbf{a}}_{\text{t}}}\left( \phi _{L} \right) \right]$, ${{\mathbf{A}}_{\text{R}}}\!=\!\left[ {{\mathbf{a}}_{\text{r}}}\left( \varphi _{1} \right)\!,\!{{\mathbf{a}}_{\text{r}}}\left( \varphi _{2} \right)\!,\!...,\!{{\mathbf{a}}_{\text{r}}}\left( \varphi _{L} \right) \right]$,\! and ${{\mathbf{H}}_{\text{a}}}\!=\!\!\sqrt{\!\frac{{{N}_{\text{t}}}{{N}_{\text{r}}}}{L}}\text{diag}({{\alpha }_{1}}\!,\!{{\alpha }_{2}},\!...,\!{{\alpha }_{L}})\!$ is an angular channel matrix \cite{CE_Ahmed_JSTSP_2014}.

	\subsection{Sparse Formulation for Signal Model}\label{S2.3}
	According to the sparsity of mmWave channels, CS theory can be used to proceed a large-scale sparse channel estimation with small-scale measurements. However, the key point is to find a suitable dictionary matrix to achieve the transformation from original channel matrix ${\bf{H}}$ to a sparse expression \cite{CSBook_Intro,CE_Ahmed_JSTSP_2014}.
	Using the quantized angle parameters proposed in \cite{CE_OMP_Korea_TCOM_2016}, the non-uniform angular grids help to construct dictionary matrices with good incoherence, which are conducive to sparse signal recovery.
	We let ${G}_{\text{t}}$ and ${G}_{\text{r}}$ denote the number of grids at BS and UE, respectively.
	The AoAs/AoDs are actually continuous in angle domain, whereas the grid quantization error is tentatively ignored here and the channel matrix can be approximated as
	\begin{align}\label{xishuxindaojuzhen}
    \mathbf{H}\approx{{\mathbf{A}}_{\text{R,G}}}{{\mathbf{H}}_{\text{a,G}}}\mathbf{A}_{\text{T,G}}^{H},
	\end{align}
	where ${{\bf{A}}_{{\text{T,G}}}} = \left[ {{{\bf{a}}_{\text{t}}}\left( {{{\bar \phi }_1}} \right),{{\bf{a}}_{\text{t}}}\left( {{{\bar \phi }_2}} \right), ... ,{{\bf{a}}_{\text{t}}}\left( {{{\bar \phi }_{G_\text{t}}}} \right)} \right] \in {\mathbb{C}^{{N_{\text{t}}} \times {G_{\text{t}}}}}$ contain ${{G}_{\text{t}}}$ array response vectors, and 
	${{\bf{A}}_{{\text{R,G}}}} = \left[ {{{\bf{a}}_{\text{r}}}\left( {{{\bar \varphi }_1}} \right),{{\bf{a}}_{\text{r}}}\left( {{{\bar \varphi }_2}} \right), ... ,{{\bf{a}}_{\text{r}}}\left( {{{\bar \varphi }_{G_\text{r}}}} \right)} \right] \in {\mathbb{C}^{{N_{\text{r}}} \times {G_{\text{r}}}}}$
	contain ${{G}_{\text{r}}}$ array response vectors. 
	Let $\bar{\phi }_{{{g}_{\text{t}}}} \in [0,\pi]$ and $\bar{\varphi }_{{{g}_{\text{r}}}} \in [0,\pi]$ denote the selected quantized angle parameters at BS and UE, respectively. Then for ${{g}_{\text{t}}}=1,2,...,{{G}_{\text{t}}}$ and ${{g}_{\text{r}}}=1,2,...,{{G}_{\text{r}}}$, we have $\cos \left( {{{\bar \phi }_{{g_t}}}} \right) = \frac{{2\left( {{g_{\rm{t}}} - 1} \right)}}{{{G_{\rm{t}}}}} - 1$ and $\cos \left( {{{\bar \varphi }_{{g_{\rm{r}}}}}} \right) = \frac{{2\left( {{g_{\rm{r}}} - 1} \right)}}{{{G_{\rm{r}}}}} - 1$.
	It can be assumed that all the elements of ${{\mathbf{H}}_{\text{a,G}}}\in {{\mathbb{C}}^{{{G}_{\text{r}}}\times {{G}_{\text{t}}}}}$ except for the $L$ items corresponding to AoAs/AoDs are zero because of the sparsity of mmWave massive MIMO channels.
	Then, by substituting (\ref{xishuxindaojuzhen}) into (\ref{hebingxingshi}), we can express the received signal at UE as
	\begin{align}\label{Y_final_form}
    \mathbf{Y}=\sqrt{P}\mathbf{W}_{\text{BB}}^{H}\mathbf{W}_{\text{RF}}^{H}
    {{\mathbf{A}}_{\text{R,G}}}{{\mathbf{H}}_{\text{a,G}}}\mathbf{A}_{\text{T,G}}^{H}
    {{\mathbf{F}}_{\text{RF}}}{{\mathbf{F}}_{\text{BB}}}+\widetilde{\mathbf{N}}.
	\end{align}

	Inspired by \cite{CE_Ahmed_JSTSP_2014,MIMOBook_Dailinglong}, to make the solving process easier, (\ref{Y_final_form}) can be vectorized as
	\begin{align}
    \nonumber {\mathbf{y}} \!=\!& \sqrt P \left( {\left( {{\bf{F}}_{{\text{BB}}}^T{\bf{F}}_{{\text{RF}}}^T} \right) \otimes \left( {{\bf{W}}_{{\text{BB}}}^H{\bf{W}}_{{\text{RF}}}^H} \right)} \right){\text{vec}}\left( {{{\bf{A}}_{{\text{R}},{\text{G}}}}{{\bf{H}}_{{\text{a}},{\text{G}}}}{\bf{A}}_{{\text{T}},{\text{G}}}^H} \right)\\
	& + {\text{vec}}({\bf{\tilde N}})\\
    \!=\!& \sqrt{\!P}\! \left( \!{\left( {{\bf{F}}_{{\text{BB}}}^T{\bf{F}}_{{\text{RF}}}^T} \right) \!\otimes\! \left( {{\bf{W}}_{{\text{BB}}}^H{\bf{W}}_{{\text{RF}}}^H} \right)} \!\right)\!\!\left( {{\bf{A}}_{{\text{T}},{\text{G}}}^ * \!\otimes\! {{\bf{A}}_{{\text{R}},{\text{G}}}}} \right)\!\!{{\bf{h}}_{{\text{a}},{\text{G}}}} \!+\! {\bf{\tilde n}}\label{CSzhankaixingshi},\!\!\!
	\end{align}
	where the identity ${\mathop{\text {vec}}\nolimits} ({\bf{ABC}}) = \left( {{{\bf{C}}^T} \otimes {\bf{A}}} \right) {\mathop{\text {vec}}\nolimits} ({\bf{B}})$ is used, and ${\mathbf{y}}=\text{vec}({\bf{Y}}) \in {\mathbb{C}^{{T_\text{t}} T_\text{r} \times 1}}$.
	The noise vector ${\bf{\tilde n}} = {\text{vec}}({\bf{\tilde N}})$, and ${{\mathbf{h}}_{\text{a,G}}}=\text{vec}({{\mathbf{H}}_{\text{a,G}}})\in {{\mathbb{C}}^{{{G}_{\text{t}}}{{G}_{\text{r}}}\times 1}}$. Besides, for simplicity, we define
	\begin{align}\label{Qzhankaixingshi}
    \mathbf{Q}\triangleq \left( {\left( {{\bf{F}}_{{\text{BB}}}^T{\bf{F}}_{{\text{RF}}}^T} \right) \otimes \left( {{\bf{W}}_{{\text{BB}}}^H{\bf{W}}_{{\text{RF}}}^H} \right)} \right)\left( {{\bf{A}}_{{\text{T}},{\text{G}}}^ *  \otimes {{\bf{A}}_{{\text{R}},{\text{G}}}}} \right),
	\end{align}
	where the dimension of $\mathbf{Q}$ is $T_\text{t} T_\text{r}\times {{G}_{\text{t}}}{{G}_{\text{r}}}$. Then, we can rewrite (\ref{CSzhankaixingshi}) as
	\begin{align}\label{biaozhunCS}
    {{\mathbf{y}}}=\sqrt{P}\mathbf{Q}{{\mathbf{h}}_{\text{a,G}}}+{{{\mathbf{\tilde{n}}}}}.
	\end{align}

	\section{Formulation for Hybrid Sensing Matrix Design}\label{S3}	

	Considering the energy overhead and computational complexity in mmWave massive MIMO systems, we generally let the number of measurements less than that of antennas (i.e. $T_\text{t}<{{N}_{\text{t}}}$ and $T_\text{r}<{{N}_{\text{r}}}$).
	In addition, for reducing performance loss caused by phase quantization while ensuring the equivalent channel matrix ${{\mathbf{H}}_{\text{a,G}}}$ is sparse enough, it is necessary to satisfy ${{G}_{\text{t}}}>{{N}_{\text{t}}}$ and ${{G}_{\text{r}}}>{{N}_{\text{r}}}$, so that we have $T_\text{t} T_\text{r}<{{G}_{\text{t}}}{{G}_{\text{r}}}$, which means that the number of rows of matrix $\mathbf{Q}$ is less than that of its columns.
	Therefore, (\ref{biaozhunCS}) is an under-determined equation, which can be solved according to the CS theory \cite{CSBook_Intro}.
	We have already transformed the channel estimation problem as a standard CS form in (\ref{CSzhankaixingshi}), where $\left( {\left( {{\bf{F}}_{{\text{BB}}}^T{\bf{F}}_{{\text{RF}}}^T} \right) \otimes \left( {{\bf{W}}_{{\text{BB}}}^H{\bf{W}}_{{\text{RF}}}^H} \right)} \right) \in {\mathbb{C}^{{T_\text{t}} T_\text{r} \times {N_{\text{t}}}{N_{\text{r}}}}}$ and $\left( \mathbf{A}_{\text{T,G}}^{\text{*}}\otimes {{\mathbf{A}}_{\text{R,G}}} \right)\in {{\mathbb{C}}^{{{N}_{\text{t}}}{{N}_{\text{r}}}\times {{G}_{\text{t}}}{{G}_{\text{r}}}}}$ correspond to the sensing matrix (alternatively called measurement matrix or projection matrix) and redundant dictionary matrix, respectively. 
	Besides, here we denote the product of $\left( {\left( {{\bf{F}}_{{\text{BB}}}^T{\bf{F}}_{{\text{RF}}}^T} \right) \otimes \left( {{\bf{W}}_{{\text{BB}}}^H{\bf{W}}_{{\text{RF}}}^H} \right)} \right)$ and $\left( \mathbf{A}_{\text{T,G}}^{\text{*}}\otimes {{\mathbf{A}}_{\text{R,G}}} \right)$ as $\mathbf{Q}$, which corresponds to the equivalent dictionary matrix in CS theory \cite{SensingOpti_SVD_ZheGong_TSP_2013,SensingOpti_AlterOptimization_ZheGong_TSP_2015}.

	For the CS-based recovery problem of sparse signal ${\mathbf{h}}_{\text{a,G}}$ in (\ref{biaozhunCS}), $\mathbf{Q}$ can be regarded as a set of ${{G}_{\text{t}}}{{G}_{\text{r}}}$ bases (alternatively called atoms) of dimension $T_\text{t} T_\text{r}\times 1$, and ${{\mathbf{h}}_{\text{a,G}}}$ can be regarded as the coefficient vector corresponding to these bases. 
	The solving process of (\ref{biaozhunCS}) is to identify the sparsest representation of ${{\mathbf{y}}}$ that uses the least number of bases/atoms. 
	The ideal situation is that the atoms of $\mathbf{Q}$ are orthogonal, but due to the dimensional constraint ($T_\text{t} T_\text{r}<{{G}_{\text{t}}}{{G}_{\text{r}}}$), it is impossible for each atom to be strictly orthogonal to each other. 
	However, the design of $\mathbf{Q}$ is still to avoid redundancy among its atoms as much as possible \cite{SensingOpti_SVD_ZheGong_TSP_2013}.

	In order to achieve a better recovery performance through designing specific equivalent dictionary matrix $\mathbf{Q}$, some conditions for $\mathbf{Q}$ have been proposed, such as restricted isometry property (RIP), exact recovery condition (ERC), and mutual incoherence property (MIP). 
	For a certain equivalent dictionary matrix $\mathbf{Q}$, it is easier for MIP to be verified compared with RIP and ERC, so MIP is generally used to design $\mathbf{Q}$ \cite{SensingOpti_CSAlgorithmAnalysis_TIT_2004,SensingOpti_OMPAnalysis_TIT_2011}. 
	The condition of MIP focuses on the mutual coherence (abbreviated as coherence in this paper) of $\mathbf{Q}$, which can be defined as
	\begin{align}\label{miu_Q_initial}
    \mu (\mathbf{Q})\triangleq \mathop {\max }\limits_{1 \le m < n \le {G_{\text{t}}}{G_{\text{r}}}} {\mkern 1mu} \frac{{\left| {{\bf{q}}_m^H{{\bf{q}}_n}} \right|}}{{{{\left\| {{{\bf{q}}_m}} \right\|}_2} {{\left\| {{{\bf{q}}_n}} \right\|}_2}}},
	\end{align}
	where ${{\mathbf{q}}_{m}}$ and ${{\mathbf{q}}_{n}}$ denote the $m$-th and $n$-th column of $\mathbf{Q}$, respectively.

	According to the identity $\left( {{\bf{AB}}} \right) \otimes \left( {{\bf{CD}}} \right) = \left( {{\bf{A}} \otimes {\bf{C}}} \right)\left( {{\bf{B}} \otimes {\bf{D}}} \right)$, we can rewrite (\ref{Qzhankaixingshi}) as
	\begin{align}\label{Q_def}
    {\bf{Q}} = \left( {{\bf{F}}_{{\text{BB}}}^T{\bf{F}}_{{\text{RF}}}^T{\bf{A}}_{{\text{T,G}}}^{\text{*}}} \right) \otimes \left( {{\bf{W}}_{{\text{BB}}}^H{\bf{W}}_{{\text{RF}}}^H{{\bf{A}}_{{\text{R,G}}}}} \right),
	\end{align}
	where we use a Kronecker product of two parts to express $\mathbf{Q}$, the two parts of ${{\bf{F}}_{{\text{BB}}}^T{\bf{F}}_{{\text{RF}}}^T{\bf{A}}_{{\text{T,G}}}^{\text{*}}}$ and ${{\bf{W}}_{{\text{BB}}}^H{\bf{W}}_{{\text{RF}}}^H{{\bf{A}}_{{\text{R,G}}}}}$ are only related to transmitting and receiving sides, respectively, so the mutual coherence in (\ref{miu_Q_initial}) can be further expressed as \cite{CSBook_Intro}
	\begin{align}\label{}
    \mu (\mathbf{Q})\!=\!\max \left\{ \mu \left( {{{\bf{F}}_{{\text{BB}}}^T{\bf{F}}_{{\text{RF}}}^T}}\mathbf{A}_{\text{T,G}}^{\text{*}} \right),\mu \left( {{{\bf{W}}_{{\text{BB}}}^H{\bf{W}}_{{\text{RF}}}^H}}{{\mathbf{A}}_{\text{R,G}}} \right) \right\}.
	\end{align}

	Thanks to the symmetry between precoding and combining matrices at transmitting and receiving sides, for a given array response matrix in (\ref{xishuxindaojuzhen}), 
	we can only focus on the hybrid combining matrix design at UE aiming for constructing a well-designed sensing matrix ${{\bf{W}}_{{\text{BB}}}^H{\bf{W}}_{{\text{RF}}}^H}$ as follows, then the sensing matrix ${{\bf{F}}_{{\text{BB}}}^T{\bf{F}}_{{\text{RF}}}^T}$ at transmitting side and equivalent dictionary matrix $\mathbf{Q}$ can be determined accordingly \cite{MIMOBook_Dailinglong,CE_Model_Korea_GCC_2014}.
	Inspired by \cite{CSBook_Intro}, proceeding with the design of hybrid combiner at UE, the optimization problem can be stated as
	\begin{subequations}
		\begin{align}
		\mathop {\text{min}} \limits_{{{\bf{W}}_{{\text{RF}}}},{{\bf{W}}_{{\text{BB}}}}} & \mu \left( {{\bf{W}}_{{\text{BB}}}^H{\bf{W}}_{{\text{RF}}}^H{\bf{A}}_{{\text{R}},{\text{G}}}} \right) \label{Coh_opt_form_a} \\
		\text{s.t.} \;\;\;\; & {{\mathbf{W}}_{\text{RF}}}(i,j)\in \mathcal{W},\forall i,j, \label{Coh_opt_cons_b} \\
		& {{\bf{W}}_{{\rm{BB}}}} \!=\! {\rm{diag}}({{\bf{W}}_{{\rm{BB,1}}}}\!,\!{{\bf{W}}_{{\rm{BB,2}}}},\!...,\!{{\bf{W}}_{{\rm{BB,}}N_{\rm{r}}^{{\rm{Block }}}}}),\! \label{Coh_opt_cons_c} \\
		& \left\| {{\mathbf{W}}_{\text{RF}}}{{\mathbf{W}}_{\text{BB}}} \right\|_F^{2}= T_\text{r}, \label{Coh_opt_cons_d}
		\end{align}
	\end{subequations}
	where the non-convex constraint in (\ref{Coh_opt_cons_b}) makes the elements of analog combining matrix $\bf{W}_\text{RF}$ satisfy constant modulus and specific phase constraints.
	The second constraint in (\ref{Coh_opt_cons_c}) ensures that all elements except those on the block-diagonal position of $\bf{W}_\text{BB}$ are equal to zeros, and the last equation in (\ref{Coh_opt_cons_d}) corresponds to the normalized power constraint.

	Here we introduce the Gram matrix $\bf{G}$ of ${{{\bf{W}}_{{\text{BB}}}^H{\bf{W}}_{{\text{RF}}}^H}}{{\mathbf{A}}_{\text{R,G}}}$ given by \cite{SensingOpti_Block_Israel_TSP_2011,SensingOpti_CSAlgorithmAnalysis_TIT_2004,SensingOpti_JointOptimization_TIP_2009}
	\begin{align}\label{Gramdefinition}
	{\bf{G}} \triangleq  {\bf{A}}_{{\rm{R,G}}}^H{{\bf{W}}_{{\rm{RF}}}}{{\bf{W}}_{{\rm{BB}}}}{\bf{W}}_{{\rm{BB}}}^H{\bf{W}}_{{\rm{RF}}}^H{{\bf{A}}_{{\rm{R,G}}}},
	\end{align}
	where ${\bf{G}} \in {{\mathbb{C}}^{{G_{\rm{r}}} \times {G_{\rm{r}}}}}$ is symmetric and positive semi-definite.
	By normalizing the columns of equivalent dictionary matrix ${{{\bf{W}}_{{\text{BB}}}^H{\bf{W}}_{{\text{RF}}}^H}}{{\mathbf{A}}_{\text{R,G}}}$ (i.e. the Frobenius norm of each column is equal to one), then we can obtain the normalized Gram matrix.
	Note that the diagonal items of normalized Gram matrix are all equal to ones, and its off-diagonal elements represent the mutual coherence between two corresponding columns of the equivalent dictionary matrix \cite{CSBook_Intro,SensingOpti_SVD_ZheGong_TSP_2013}.

	In fact, mutual coherence $\mu \left( {{\bf{W}}_{{\text{BB}}}^H{\bf{W}}_{{\text{RF}}}^H{\bf{A}}_{{\text{R}},{\text{G}}}} \right)$ in (\ref{Coh_opt_form_a}) is just a worst-case value and cannot reflect the average signal recovery performance \cite{SensingOpti_OMPAnalysis_TIT_2011}. 
	Besides, mutual coherence is also difficult to be optimized directly in practical application. 	
	Therefore, many researchers have made a slight approximation and proposed to minimize the Frobenius norm of the difference between Gram matrix and identity matrix, which can minimize the off-diagonal elements of the Gram matrix and decrease the mutual coherence indirectly.
	Thus, the design of hybrid combiner at UE can be further formulated as\cite{CE_Model_Korea_GCC_2014,CE_OMP_Korea_TCOM_2016}
	\begin{subequations} \label{New_Coh_opt_tot}
		\begin{align}
			\mathop {\text{min}}\limits_{{{\bf{W}}_{{\text{RF}}}},{{\bf{W}}_{{\text{BB}}}}} & {\left\| {{\bf{A}}_{{\rm{R,G}}}^H{{\bf{W}}_{{\rm{RF}}}}{{\bf{W}}_{{\rm{BB}}}}{\bf{W}}_{{\rm{BB}}}^H{\bf{W}}_{{\rm{RF}}}^H{{\bf{A}}_{{\rm{R,G}}}} \!-\! {{\bf{I}}_{{G_\text{r}}}}} \right\|}_F^2 \label{New_Coh_opt_form_a} \\
			\text{s.t.} \;\;\;\; & {{\mathbf{W}}_{\text{RF}}}(i,j)\in \mathcal{W},\forall i,j, \label{New_Coh_opt_cons_b} \\
			& {{\bf{W}}_{{\rm{BB}}}} \!=\! {\rm{diag}}({{\bf{W}}_{{\rm{BB,1}}}}\!,\!{{\bf{W}}_{{\rm{BB,2}}}},\!...,\!\!{{\bf{W}}_{{\rm{BB,}}N_{\rm{r}}^{{\rm{Block }}}}}) \label{New_Coh_opt_cons_c},
		\end{align}
	\end{subequations}
	where the constraints in (\ref{Coh_opt_cons_b}) and (\ref{Coh_opt_cons_c}) are remained. 
	The power constraint in (\ref{Coh_opt_cons_d}) is removed, because multiplying a coefficient on $\bf{W}_\text{BB}$ will not change the coherence of any two columns of ${{{\bf{W}}_{{\text{BB}}}^H{\bf{W}}_{{\text{RF}}}^H}}{{\mathbf{A}}_{\text{R,G}}}$ \cite{CSBook_Intro}.
	Therefore, we can temporarily ignore (\ref{Coh_opt_cons_d}), and calculate the proper coefficient multiplied on $\bf{W}_\text{BB}$ according to the power constraint after solving (\ref{New_Coh_opt_tot}).

	\section{Hybrid Sensing Matrix Design with Infinite-Resolution PSs}\label{S4}

	The hybrid analog-digital sensing matrix design problem (\ref{New_Coh_opt_tot}) not only faces the constant modulus constraints, but also the discrete phase constraints.
	In this section, we first focus on the scenario where the resolution of PSs is infinite, which can be considered as an extreme case of low-resolution PSs mentioned later.
	Thus, problem (\ref{New_Coh_opt_tot}) can be simplified as
	\begin{subequations} \label{Infreso_Coh_opt_tot}
		\begin{align}
			\mathop {\text{min}}\limits_{{{\bf{W}}_{{\text{RF}}}},{{\bf{W}}_{{\text{BB}}}}} & {\left\| {{\bf{A}}_{{\rm{R,G}}}^H{{\bf{W}}_{{\rm{RF}}}}{{\bf{W}}_{{\rm{BB}}}}{\bf{W}}_{{\rm{BB}}}^H{\bf{W}}_{{\rm{RF}}}^H{{\bf{A}}_{{\rm{R,G}}}} - {{\bf{I}}_{{G_\text{r}}}}} \right\|}_F ^2 \label{Infreso_Coh_opt_form_a} \\
			\text{s.t.} \;\;\;\; & \left| {{\mathbf{W}}_{\text{RF}}}(i,j) \right| =1,\forall i,j, \label{Infreso_Coh_opt_cons_b} \\
			& {{\bf{W}}_{{\rm{BB}}}} \!=\! {\rm{diag}}({{\bf{W}}_{{\rm{BB,1}}}}\!,\!{{\bf{W}}_{{\rm{BB,2}}}},\!...,\!{{\bf{W}}_{{\rm{BB,}}N_{\rm{r}}^{{\rm{Block }}}}}) \label{Infreso_Coh_opt_cons_c}.\!
		\end{align}
	\end{subequations}

	Considering the non-convexity of (\ref{Infreso_Coh_opt_tot}), here we propose the following iterative algorithm to obtain a sub-optimal solution. 
	Since the analog and digital combiner face different types of constraints, it is difficult to proceed an effective joint optimization. 
	Therefore, we approximate the optimal solution by iteratively optimizing one of them when the other is fixed. 
	
	\subsection{Digital Combiner Design}\label{S4.1}

	We first consider the design of $\bf{W}_\text{BB}$ when $\bf{W}_\text{RF}$ is fixed, which can be expressed as
	\begin{subequations} \label{OptBB_Coh_opt_tot}
		\begin{align}
			\mathop {\text{min}}\limits_{{{\bf{W}}_{{\text{BB}}}}} \;\; & 
			{\left\|{ \! {{\bf{A}}_{{\rm{R,G}}}^H{{\bf{W}}_{{\rm{RF}}}}{{\bf{W}}_{{\rm{BB}}}}{\bf{W}}_{{\rm{BB}}}^H{\bf{W}}_{{\rm{RF}}}^H{{\bf{A}}_{{\rm{R,G}}}} - {{\bf{I}}_{{G_\text{r}}}}} \!}\right\|}_F ^2 \label{OptBB_Coh_opt_form_a} \\
			\text{s.t.} \;\;\; & {{\bf{W}}_{{\rm{BB}}}} = {\rm{diag}}({{\bf{W}}_{{\rm{BB,1}}}},{{\bf{W}}_{{\rm{BB,2}}}},...,{{\bf{W}}_{{\rm{BB,}}N_{\rm{r}}^{{\rm{Block }}}}}) \label{OptBB_Coh_opt_cons_b},
		\end{align}
	\end{subequations}
	where the Frobenius norm of the quadratic term of $\bf{W}_\text{BB}$ makes it non-convex and difficult to be solved directly, so we introduce an intermediate variable $\bf{X}_\text{BB}$ denoted as
	\begin{align}\label{X_BB}
		{{\bf{X}}_{{\rm{BB}}}} = {\rm{diag}}({{\bf{X}}_{{\rm{BB,1}}}},{{\bf{X}}_{{\rm{BB,2}}}},...,{{\bf{X}}_{{\rm{BB,}}N_{\rm{r}}^{{\rm{Block }}}}}),
	\end{align}
	where ${{\bf{X}}_{{\rm{BB,}}q}} = {{\bf{W}}_{{\rm{BB,}}q}}{\bf{W}}_{{\rm{BB,}}q}^H$ is a positive semi-definite matrix with dimension of ${N^{{\rm{RF}}}} \times {N^{{\rm{RF}}}}$, for $q = 1,2, ... ,N_{\rm{r}}^{{\rm{Block }}}$.
	\begin{figure*}[!t]
		\normalsize
		\begin{subequations} \label{OptBBblock_Coh_opt_tot}
			\begin{align}
				\mathop {\text{min}}\limits_{{\bf{X}}_{\text{BB},1},{{\bf{X}}}_{\text{BB},2},...,{\bf{X}}_{\text{BB},{N}_\text{r}^\text{Block}}} & 
				{\left\| {{\bf{A}}_{{\rm{R,G}}}^H{{{\bf{W}}}_{{\rm{RF}}}}{\text{diag}({{\bf{X}}_{\text{BB},1},{\bf{X}}_{\text{BB},2},...,{\bf{X}}_{\text{BB},N_\text{r}^\text{Block}}})}{\bf{W}}_{{\rm{RF}}}^H{{\bf{A}}_{{\rm{R,G}}}} - {{\bf{I}}_{{G_\text{r}}}}} \right\|}_F ^2 \label{OptBBblock_Coh_opt_form_a} \\
				\text{s.t. } \;\;\;\;\;\;\;\;\;\;\;\; & {{\bf{X}}_{{\text{BB,}}q}} \succeq 0,\forall 1\leq q\leq N_\text{r}^\text{Block} \label{OptBBblock_Coh_opt_cons_b}
			\end{align}
		\end{subequations}
		\hrulefill
		\vspace*{4pt}
	\end{figure*}
	Thus, by rewriting (\ref{OptBB_Coh_opt_tot}), (\ref{OptBBblock_Coh_opt_tot}) has already been transformed into a convex expression, hence we can solve it by several approaches, such as the generic CVX solver.
	Then, ${{\bf{W}}_{{\text{BB,}}q}}$ can be given by 
	\begin{align}
		{{\bf{W}}_{{\rm{BB,}}q}} = {{\hat{\bf{V}}}_{{\rm{X}},q}}{\hat{\bf{\Sigma }}}_{{\rm{X}},q}^{{\textstyle{1 \over 2}}}, \label{EigenDecom}
	\end{align}
	where ${\hat{\bf{V}}}_{{\text{X}},q} \in \mathbb{C}^{N_\text{RF} \times N_\text{s}}$ and ${\hat{\bf{\Sigma}}}_{{\text{X}},q} \in \mathbb{C}^{N_\text{s} \times N_\text{s}}$ denote the matrices composed of the $N_\text{s}$ strongest eigenvectors and eigenvalues of ${{\bf{X}}_{{\text{BB,}}q}}$, respectively, and $\bf{W}_\text{BB}$ can be obtained by superposing ${{\mathbf{W}}_{\text{BB},1}},...,$ ${{\mathbf{W}}_{\text{BB},N_{\text{r}}^{\text{Block}}}}$ on its diagonal. 

	\subsection{Analog Combiner Design}\label{S4.2}
	
	In this subsection, we turn to design the analog combiner $\bf{W}_\text{RF}$ with a fixed digital combiner $\bf{W}_\text{BB}$, when the resolution of PSs is unlimited, it can be expressed as
	\begin{subequations} \label{OptRF_Coh_opt_tot}
		\begin{align}
			\mathop {\text{min}}\limits_{{{\bf{W}}_{{\text{RF}}}}} & \;
			{\left\| {{\bf{A}}_{{\rm{R,G}}}^H{{\bf{W}}_{{\rm{RF}}}}{{\bf{W}}_{{\rm{BB}}}}{\bf{W}}_{{\rm{BB}}}^H{\bf{W}}_{{\rm{RF}}}^H{{\bf{A}}_{{\rm{R,G}}}} - {{\bf{I}}_{{G_\text{r}}}}} \right\|}_F ^2 \label{OptRF_Coh_opt_form_a} \\
			\text{s.t.} \;&\; \left| {{\mathbf{W}}_{\text{RF}}}(i,j) \right| =1,\forall i,j \label{OptRF_Coh_opt_cons_b},
		\end{align}
	\end{subequations}
	where the non-convexity of (\ref{OptRF_Coh_opt_cons_b}) brings great challenges to solve (\ref{OptRF_Coh_opt_tot}) optimally.
	Fortunately, although the constraint in (\ref{OptRF_Coh_opt_cons_b}) is non-convex, it is smooth. 
	Thus, we can consider (\ref{OptRF_Coh_opt_cons_b}) as a specific type of domain \cite{ManifoldOpti_AltMin_JSTSP_2016}, then (\ref{OptRF_Coh_opt_tot}) turns to be an unconstrained optimization problem in a nonlinear space according to the manifold optimization theory. 
	We introduce some definitions and terminologies of manifold optimization, 
	and more details can be found in \cite{ManiOptBook_1_Optimization_Algorithms_on_Matrix_Manifolds,ManiOptBook_2_Manifold_Learning_Theory_and_Applications}.
	\begin{figure}[t]
		\center{\includegraphics[width=0.85\columnwidth]{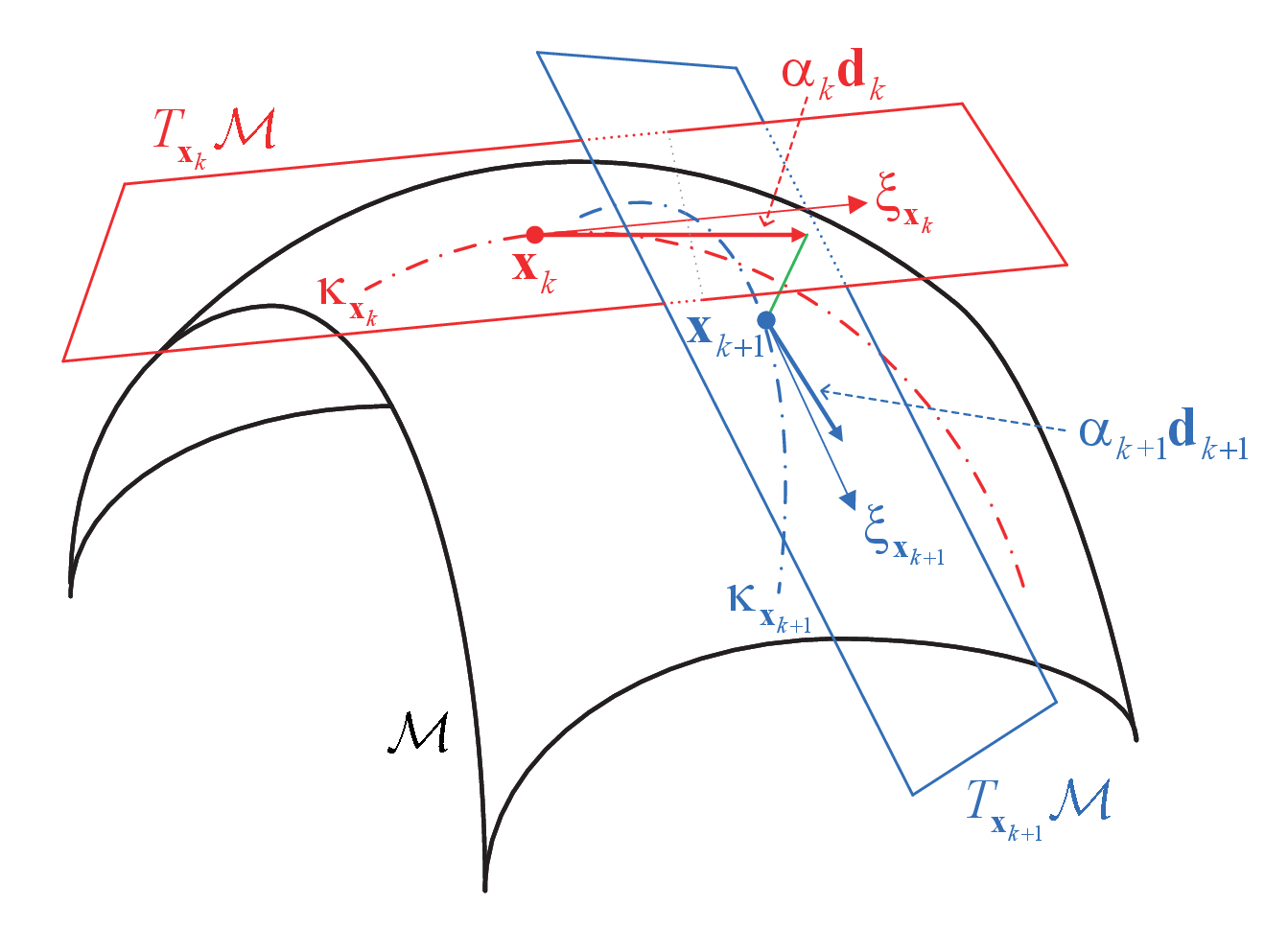}}
		\caption{The process of iterative optimization on a Riemannian manifold.}
		\label{fig2}
	\end{figure}

	In order to express this more intuitively, the process of iterative optimization on a manifold is shown in Fig. \ref{fig2}, where the subscripts $k$ and $k+1$ denote the variables in the $k$-th and $(k+1)$-th iteration, respectively.
	The tangent space denoted as ${T_{{{\bf{x}}}}}{\cal M}$ at a given point ${\bf{x}}$ on the manifold ${\cal M}$ is composed of the tangent vectors ${\xi _{{\bf{x}}}}$ of the curves ${\kappa _{{\bf{x}}}}$ through ${\bf{x}}$ \cite{ManiOptBook_1_Optimization_Algorithms_on_Matrix_Manifolds,ManifoldOpti_AltMin_JSTSP_2016}.
	Inspired by \cite{ManiOpt_applyinComm_Gangke_ISIT_2015,ManifoldOpti_AltMin_JSTSP_2016}, we can vectorize $\bf{W}_\text{RF}$ and obtain the constant modulus vector $\bf{x}$ with dimension of $N_\text{r}{M_\text{r}} \times 1$, which forms a complex circle manifold (called Riemannian manifold) expressed as
	\begin{align}\label{manifold_ini}
		{\cal M}_{cc}^{{N_{\rm{r}}}{M_{{\rm{r}}}}} = \left\{ {{\bf{x}}:\left| {\bf{x}} \right| = {{\bf{1}}_{{N_{\rm{r}}}{M_{{\rm{r}}}} , 1}}} \right\},
	\end{align}
	and (\ref{manifold_ini}) can also be equivalently written as
	\begin{align}\label{manifold_change}
		{\cal M}_{cc}^{{N_{\rm{r}}}{M_{{\rm{r}}}}} = \left\{ {{\bf{x}}:\Re \left\{ {{{\bf{x}}^*} \circ {\bf{x}}} \right\} - {{\bf{1}}_{{N_{\rm{r}}}{M_{{\rm{r}}}} , 1}} = {{\bf{0}}_{{N_{\rm{r}}}{M_{{\rm{r}}}} , 1}}} \right\},
	\end{align}	
	where the operator $\circ $ represents the Hadamard product.
	
	In Euclidean space (linear space), the gradient-based method is always used to solve matrix optimization problems, but it cannot be directly applied to aforementioned problems, because the obtained solution may not be on the Riemannian manifold after updating along the direction of Euclidean gradient.
	Therefore, for a given point $\bf{x}$ on the manifold of (\ref{manifold_change}), we try to keep it still on the Riemannian manifold after updating along a certain vector ${\bf{z}}\in{\mathbb{C}^{{N_{\text{r}}}{M_{{\rm{r}}}} \times 1}}$ according to the manifold optimization theory, thus we have
	\begin{align}\label{Tan_space_procedure}
		\nonumber & \text{ } \Re \left\{ {\left( {{{\bf{x}}^*} + {{\bf{z}}^*}} \right) \circ \left( {{\bf{x}} + {\bf{z}}} \right)} \right\} - {{\bf{1}}_{{N_{\rm{r}}}{M_{{\rm{r}}}} , 1}} \\
		\nonumber = & \text{ } \Re \left\{ {{{\bf{x}}^*} \circ {\bf{x}}} \right\} - {{\bf{1}}_{{N_{\rm{r}}}{M_{{\rm{r}}}} , 1}} + 2\Re \left\{ {{{\bf{x}}^*} \circ {\bf{z}}} \right\} + \Re \left\{ {{{\bf{z}}^*} \circ {\bf{z}}} \right\} \\
		= & \text{ } {{\bf{0}}_{{N_{\rm{r}}}{M_{{\rm{r}}}} , 1}},
	\end{align}	
	where ${\Re \left\{ {{{\bf{x}}^*} \circ {\bf{x}}} \right\} - {{\bf{1}}_{{N_{\rm{r}}}{M_{{\rm{r}}}} , 1}} = {{\bf{0}}_{{N_{\rm{r}}}{M_{{\rm{r}}}} , 1}}}$, because $\bf{x}$ is on the Riemannian manifold. 
	Besides, the update stepsize is usually extremely small, which means $\Re \left\{ {{{\bf{z}}^*} \circ {\bf{z}}} \right\} \approx  {{\bf{0}}_{{N_{\rm{r}}}{M_{{\rm{r}}}} , 1}}$, so the tangent space at a given point $\bf{x}$ is composed of all possible $\bf{z}$, which can be given by
	\begin{align}\label{Tan_space}
		{T_{\bf{x}}}{\cal M}_{cc}^{{N_{\rm{r}}}{M_{{\rm{r}}}}} = \left\{ {{\bf{z}}:\Re \left\{ {{{\bf{x}}^*} \circ {\bf{z}}} \right\} = {{\bf{0}}^{{N_{\rm{r}}}{M_{{\rm{r}}}} \times 1}}} \right\}.
	\end{align}	
	 
	Based on (\ref{OptRF_Coh_opt_form_a}), here we define the cost function as
	\begin{align}\label{costfuc_infreso_RF}
		f({\bf{x}}) \! \buildrel \Delta \over = \!\! \left\| {{\bf{A}}_{{\rm{R}},{\rm{G}}}^H{\mathop{\rm invec\!}\nolimits} \left( {\bf{x}} \right)\!{{\bf{W}}_{{\rm{BB}}}}\!{\bf{W}}_{{\rm{BB}}}^H{{\left( {{\mathop{\rm invec\!}\nolimits} \left( {\bf{x}} \right)} \right)}^H}\!\!{{\bf{A}}_{{\rm{R}},{\rm{G}}}} \!-\! {{\bf{I}}_{{G_{\rm{r}}}}}} \right\|_F^2 \!,
	\end{align}		
	where ${{\mathop{\rm invec}\nolimits} \left( {\bf{x}} \right)} \in {\mathbb{C}^{{N_{\rm{r}}} \times {M_{{\rm{r}}}}}}$ corresponds to the analog combiner, then its Euclidean gradient can be formulated as
	\begin{align}\label{Euc_Gra}
		\nonumber  \nabla\! f({\bf{x}}) \!\!= & 4{\rm{vec\!}}\left[ {{{\bf{A}}_{{\rm{R}},{\rm{G}}}}{\bf{A}}_{{\rm{R}},{\rm{G}}}^H{\mathop{\rm invec}\nolimits} ({\rm{x}}){{\bf{W}}_{{\rm{BB}}}}{\bf{W}}_{{\rm{BB}}}^H} \right. \\
		& \!\times\! \left.\!\!{\left(\!\! {{{({\mathop{\rm invec}\nolimits} ({\bf{x}}))}^H}{\!\!{\bf{A}}_{{\rm{R}},{\rm{G}}}}{\bf{A}}_{{\rm{R}},{\rm{G}}}^{\!H}{\mathop{\rm invec}\nolimits} ({\bf{x}}){{\bf{W}}_{{\rm{BB}}}}\!{\bf{W}}_{{\rm{BB}}}^H \!-\! {{\bf{I}}_{{M_{\rm{r}}}}}} \!\right)} \!\right]\!.\!
	\end{align}
	To avoid the update of $\bf{x}$ deviating from the manifold as much as possible, we can obtain the Riemannian gradient (denoted as $\rm{grad} (\cdot)$) by projecting the Euclidean gradient $\nabla f({\bf{x}})$ onto the tangent space ${T_{\bf{x}}}{\cal M}_{cc}^{{N_{\rm{r}}}{M_{{\rm{r}}}}}$ orthogonally \cite{ManifoldOpti_AltMin_JSTSP_2016}, which can be given by 
	\begin{align}\label{Rie_Gra}
		{\mathop{\rm grad}\nolimits} f({\bf{x}}) \!=\! {{\mathop{\rm \!Proj}\nolimits} _{\bf{x}}}\nabla f({\bf{x}}) \!=\! \nabla f({\bf{x}}) \!-\! \Re \!\left\{ {\nabla f({\bf{x}}) \!\circ\! {{\bf{x}}^*}} \right\} \!\circ\! {\bf{x}}.\!
	\end{align}
	
	Although we assume the updating vector $\bf{z}$ in (\ref{Tan_space_procedure}) is extremely small, it is not exactly $\bf{0}$ in fact, which makes the obtained point slightly deviate from the Riemannian manifold.
	Therefore, we introduce another important terminology `retraction' (denoted as ${{\mathop{\rm Retr}\nolimits} _{\bf{x}}}( \cdot )$) to map the points deviated from the manifold after updating along the conjugate direction $\bf{d}$ onto the Riemannian manifold again, which can be given by
	\begin{align}\label{Retrac}
		\nonumber {{\mathop{\rm Retr}\nolimits} _{\bf{x}}}: \text{ } & {T_{\bf{x}}}{\cal M}_{cc}^{{N_{\rm{r}}}{M_{{\rm{r}}}}} \to {\cal M}_{cc}^{{N_{\rm{r}}}{M_{{\rm{r}}}}} \\
		 & \alpha {\bf{d}} \mapsto {{\mathop{\rm Retr}\nolimits} _{\bf{x}}}(\alpha {\bf{d}}) = {\mathop{\rm vec}\nolimits} \left[ {\frac{{{{({\bf{x}} + \alpha {\bf{d}})}_i}}}{{\left| {{{({\bf{x}} + \alpha {\bf{d}})}_i}} \right|}}} \right],
	\end{align}
	where ${{{({\bf{x}} + \alpha {\bf{d}})}_i}}$ denotes the $i$-th element of vector ${({\bf{x}} + \alpha {\bf{d}})}$, for $i = 1,2,...,{N_{\rm{r}}}{M_{{\rm{r}}}}$, and $\alpha$ is the stepsize multiplied on $\bf{d}$.	
	As shown in Fig. \ref{fig2}, ${\bf{x}}_k$ has deviated from ${\cal M}$ after updating along ${\alpha _k}{{\bf{d}}_k}$, then we can remap it onto the Riemannian manifold at ${\bf{x}}_{k+1}$ through retraction.
	\begin{algorithm}[th!]
		\renewcommand{\algorithmicrequire}{\textbf{Input:}}
		\renewcommand\algorithmicensure {\textbf{Output:} }
		\caption{Proposed Riemannian Manifold Gradient Descent based Algorithm for Analog combiner Design (inspired by \cite{ManifoldOpti_AltMin_JSTSP_2016})}
		\label{}
		\begin{algorithmic}[1]
			\STATE\textbf{Input:} ${{{\bf{A}}_{{\rm{R,G}}}}}$, ${\bf{W}}_\text{BB}$;
			\STATE\textbf{Initialization:}
			Construct $\bf{W}_\text{\!RF,0}$ with random phases,\! $k\!=\!0$; \\
			\STATE ${{\bf{x}}_0} = {\rm{vec}}\left( {{{\bf{W}}_{{\rm{RF}},0}}} \right)$ and ${{\bf{d}}_0} =  - {\mathop{\rm grad}\nolimits} f({{\bf{x}}_0})$; \\
			\REPEAT
			\STATE Choose Armijo backtracking line search step size ${\alpha _k}$;
			\STATE Obtain ${{\bf{x}}_{k + 1}} = {{\mathop{\rm Retr}\nolimits} _{\bf{x}}}({\alpha _k}{{\bf{d}}_k})$ by (\ref{Retrac});
			\STATE Calculate the Euclidean gradient $\nabla f({\bf{x}}_{k+1})$ by (\ref{Euc_Gra});
			\STATE Calculate the Riemannian gradient ${\mathop{\rm grad}\nolimits} f({{\bf{x}}_{k+1}})$ by (\ref{Rie_Gra});
			\STATE Calculate the Polak-Ribiere parameter ${\beta _{k + 1}}$ by (\ref{PR_exp}); \label{PR_cal}
			\STATE Obtain ${\bf{d}}_k^ + $ by transporting ${{\bf{d}}_k}$ from ${{\bf{x}}_k}$ to ${{\bf{x}}_{k + 1}}$ according to (\ref{Transport});
			\STATE ${{\bf{d}}_{k + 1}} \!=\!  - {\mathop{\rm grad}\nolimits} f({{\bf{x}}_{k + 1}}) \!+\! {\beta _{k + 1}}{\bf{d}}_k^ + $; \!\%\! conjugate direction
			\STATE $k \leftarrow k + 1$;
			\UNTIL a stopping criterion triggers.
			\STATE Obtain the final analog combiner ${{\bf{W}}_{{\rm{RF}}}} = {\rm{invec}}\left( {{{\bf{x}}_{k - 1}}} \right)$;
			\STATE\textbf{Output:} ${{\bf{W}}_{{\rm{RF}}}}$ \\
		\end{algorithmic}
	\end{algorithm}

	According to literatures \cite{ManiOptBook_1_Optimization_Algorithms_on_Matrix_Manifolds,ManiOptBook_2_Manifold_Learning_Theory_and_Applications}, it is necessary to slightly correct the updating direction. 
	Therefore, $\bf{d}$ and ${\xi _{\bf{x}}}$ are not collinear, and $\bf{d}$ is not on the tangent space ${T_{\bf{x}}}{\cal M}$ strictly as shown in Fig. \ref{fig2}.
	The correction term is related to the updating vector in the last iteration, so we introduce the terminology `Transport', which can map the conjugate direction ${\bf{d}}_k$ in the previous $k$-th iteration onto the updated tangent space at ${\bf{x}}_{k+1}$, thereby correcting the new conjugate direction ${\bf{d}}_{k+1}$. 
	The correcting direction (denoted as ${\bf{d}}_k^ + $) obtained by `Transport' can be expressed as
	\begin{align}\label{Transport}
		\nonumber {{\mathop{\rm \!Transp}\nolimits} _{{{\bf{x}}_k} \to {{\bf{x}}_{k \!+\! 1}}}}\!:\;& {T_{{{\bf{x}}_k}}}{\cal M}_{cc}^{{N_{\rm{r}}}{M_{{\rm{r}}}}} \to {T_{{{\bf{x}}_{k + 1}}}}{\cal M}_{cc}^{{N_{\rm{r}}}{M_{{\rm{r}}}}} \\
		& {{\bf{d}}_k} \mapsto {\bf{d}}_k^ +  \!=\! {{\bf{d}}_k} \!- \Re\! \left\{ {{{\bf{d}}_k} \!\circ\! {\bf{x}}_{k +\! 1}^*} \right\} \!\circ\! {{\bf{x}}_{k +\! 1}}.
	\end{align}	

	Inspired by \cite{ManifoldOpti_AltMin_JSTSP_2016}, here we propose a Riemannian manifold gradient descent based algorithm for analog combiner design according to aforementioned steps and the classical line search based conjugate gradient method \cite{ManiOptBook_1_Optimization_Algorithms_on_Matrix_Manifolds}, which can be summarized as Algorithm 1.
	The coefficient multiplied on ${\bf{d}}_k^ + $ is closely related to the convergence \cite{ManiOptBook_1_Optimization_Algorithms_on_Matrix_Manifolds}, so for keeping the objective function in (\ref{OptRF_Coh_opt_form_a}) non-increasing in each iteration, Armijo backtracking line search step and conjugate direction Polak-Ribiere parameter are introduced in Algorithm 1, which are described in detail in \cite{ManiOptBook_1_Optimization_Algorithms_on_Matrix_Manifolds,ManiOptBook_2_Manifold_Learning_Theory_and_Applications}.
	The well known Polak-Ribiere parameter ${\beta _{k + 1}}$ in step (\ref{PR_cal}) can be given by
	\begin{align}\label{PR_exp}
		{\beta _{k + 1}} \!=\! \frac{{{{\left( {{\mathop{\rm grad}\nolimits} f\left( {{{\bf{x}}_{k + 1}}} \right)} \right)}^H}\!\!\left( {{\mathop{\rm grad}\nolimits} f\left( {{{\bf{x}}_{k + 1}}} \right) \!-\! {\mathop{\rm grad}\nolimits} f\left( {{{\bf{x}}_k}} \right)} \right)}}{{{{\left( {{\mathop{\rm grad}\nolimits} f\left( {{{\bf{x}}_k}} \right)} \right)}^H}{\mathop{\rm grad}\nolimits} f\left( {{{\bf{x}}_k}} \right)}}.
	\end{align}	

	The convergence of Algorithm 1 is guaranteed according to Theorem 4.3.1 in \cite{ManiOptBook_1_Optimization_Algorithms_on_Matrix_Manifolds}, and the iteration stops when the cost function (\ref{costfuc_infreso_RF}) no longer decreases \cite{ManifoldOpti_AltMin_JSTSP_2016}.

	\subsection{Hybrid Combiner Design}\label{S4.3}

	After solving the optimization problem of analog/digital combiner when the other is fixed, a hybrid combiner design algorithm with infinite-resolution PSs is proposed, which is described in Algorithm 2.
	Step \ref{Power_cons} calculates the coefficient multiplied on digital combiner ${{\bf{W}}_{{\rm{BB}}}}$ according to the power constraint, and the iteration stops when the value of (\ref{Infreso_Coh_opt_form_a}) no longer decreases.

	\begin{algorithm}[th!]
		\renewcommand{\algorithmicrequire}{\textbf{Input:}}
		\renewcommand\algorithmicensure {\textbf{Output:} }
		\caption{Proposed Alternating Hybrid Analog-Digital Combiner Design Algorithm with Infinite-Resolution PSs}
		\label{}
		\begin{algorithmic}[1]
			\STATE\textbf{Input:} ${{{\bf{A}}_{{\rm{R,G}}}}}$;
			\STATE\textbf{Initialization:}
			Construct $\bf{W}_\text{\!RF,0}$ with random phases,\! $k\!=\!0$; \\
			\REPEAT
			\STATE Optimize ${{\bf{W}}_{{\text{BB,}}k+1}}$ with fixed ${{\bf{W}}_{{\text{RF,}}k}}$ by solving (\ref{OptBB_Coh_opt_tot});
			\STATE Fix ${{\bf{W}}_{{\text{BB,}}k+1}}$, obtain ${{\bf{W}}_{{\text{RF,}}k+1}}$ by Algorithm 1;
			\STATE $k \leftarrow k + 1$ ;
			\UNTIL a stopping criterion triggers.			
			\STATE Obtain the final hybrid combiner ${{\bf{W}}_{{\rm{RF}}}} ={{\bf{W}}_{{\rm{RF}},k-1}}$ and ${{\bf{W}}_{{\rm{BB}}}} = \frac{{\sqrt {{T_\text{r}}} {{\bf{W}}_{{\rm{BB,}}k - 1}}}}{{{{\left\| {{{\bf{W}}_{{\rm{RF}}}}{{\bf{W}}_{{\rm{BB,}}k - 1}}} \right\|}_F}}}$; \label{Power_cons}
			\STATE\textbf{Output:} ${{\bf{W}}_{{\rm{RF}}}}$, ${{\bf{W}}_{{\rm{BB}}}}$ \\
		\end{algorithmic}
	\end{algorithm}	

	We proceed with our complexity analysis of the proposed Algorithm 2 for infinite-resolution PSs scenario as follows. 
	The computational complexity is composed of two parts, namely the digital combiner design and the analog combiner design.
	As for the digital combiner design, the positive semi-definite matrix ${{\bf{X}}_{{\rm{BB}}}} \in \mathbb{C}^{M_\text{r} \times M_\text{r}}$ in (\ref{X_BB}) is obtained using the generic CVX solver at a computational complexity order of $\mathcal{O}\left(M_{\text {r}}^3\right)$ for each iteration.
	Thus, the design of digital combiner requires about $\mathcal{O}\left(N_\text{iter,1}^\text{d} M_{\text {r}}^3\right)$ computations, where $N_\text{iter,1}^\text{d}$ denotes the number of iterations for CVX.
	The updating of analog combiner needs $\mathcal{O}\left(N_\text{r} G_\text{r}^2\right)$ computations for each iteration with an appropriate fixed stepsize.
	Then the computational complexity of analog combiner design is $\mathcal{O}\left(N_\text{iter,1}^\text{a} N_\text{r} G_\text{r}^2\right)$, where $N_\text{iter,1}^\text{a}$ denotes the number of outer iterations in Algorithm 1.
	Therefore, the overall computational complexity of Algorithm 2 can be approximately expressed as $\mathcal{O}\left(N_\text{iter}^\text{h} (N_\text{iter,1}^\text{d} M_{\text {r}}^3 + N_\text{iter,1}^\text{a} N_\text{r} G_\text{r}^2)\right)$, where $N_\text{iter}^\text{h}$ denotes the number of alternating analog-digital iterations.

	\section{Hybrid Sensing Matrix Design with Low-Resolution PSs}\label{S5}

	We propose the design of the hybrid combiner with infinite-resolution PSs in the last section.
	However, considering the hardware cost and power consumption, the resolution of PSs is limited in actual applications \cite{Background_Overview_Heath_JSTSP_2016,Background_RFswitch_Heath_Access_2016,LowPSPrecoding_YuWei_JSTSP_2015}.
	Therefore, we will design the hybrid sensing matrix with low-resolution (1-3 bits) PSs in this section.

	For the low-resolution PSs scenario, a straightforward idea is to apply Algorithm 1 and then quantize the phases to obtain $\bf{W}_\text{RF}$, and $\bf{W}_\text{BB}$ can be optimized using the CVX solver according to (\ref{OptBB_Coh_opt_tot}-\ref{EigenDecom}).
	However, during the alternating iterations of the aforementioned two steps, a severe performance degradation may occur.
	The main reason is that the phase quantization step will inevitably cause the performance loss. When the resolution of PSs is extremely low (such as 1,2,3 bits), the performance loss brought by the direct quantization of the whole $\bf{W}_\text{RF}$ is particularly severe, which may even make the alternating iterations fail to converge.
	Therefore, we adopt a relatively mild strategy, which approximates the identity matrix by the sum of $N_\text{r}^\text{Block}$ Gram submatrices instead of the whole Gram matrix as 
	\begin{align}\label{block-wise_initial_obj}
		{\left\| {\sum\limits_{q = 1}^{N_{\rm{r}}^{{\rm{Block}}}} \!{\!{\bf{A}}_{{\rm{R,G}}}^H{{\bf{W}}_{{\rm{RF\!,}}q}}{{\bf{W}}_{{\rm{BB\!,}}q}}{\bf{W}}_{{\rm{BB\!,}}q}^H{\bf{W}}_{{\rm{RF\!,}}q}^H{{\bf{A}}_{{\rm{R,G}}}}} \!-\! {{\bf{I}}_{{G_\text{r}}}}} \right\|_F ^2}\!.\!
	\end{align}	
	It means that we can equivalently divide the entire optimization problem into multiple sub-problems.
	Furthermore, the optimization of analog combiner ${{\bf{W}}_{{\rm{RF,}}q}}$ and digital combiner ${{\bf{W}}_{{\rm{BB,}}q}}$ in the $q$-th block with the ones in other blocks fixed can be equivalently expressed as
	\begin{subequations} \label{Blk_opt}
		\begin{align}
			\mathop {\text{min}}\limits_{{{\bf{W}}_{{\text{RF}}\!,q}}\!,\!{{\bf{W}}_{{\text{BB}}\!,q}}} & 
			{\left\| {{\bf{A}}_{{\rm{R,\!G}}}^H\!{{\bf{W}}_{{\rm{RF\!,}}q}}\!{{\bf{W}}_{{\rm{BB\!,}}q}}\!{\bf{W}}_{{\rm{BB\!,}}q}^H\!{\bf{W}}_{{\rm{RF\!,}}q}^H{{\bf{A}}_{{\rm{R,\!G}}}} \!-\! {{\bf{E}}_q}} \right\|_F ^2} \label{Blk_opt_form_a} \\
			\text{s.t.} \;\;\;\;\; & \; {{\mathbf{W}}_{{\text{RF}},q}}(i,j)\in \mathcal{W},\forall i,j \label{Blk_opt_cons_b},
		\end{align}
	\end{subequations}
	where the Hermittian matrix
	\begin{align}\label{E_q}
		{{\bf{E}}_q} \!=\! {{\bf{I}}_{{G_\text{r}}}} \!-\! \sum\limits_{t = 1\!,t \ne q}^{N_{\rm{r}}^{{\rm{Block }}}} \!{{\bf{A}}_{{\rm{R,G}}}^H\!{{\bf{W}}_{{\rm{\!RF\!,}}t}}\!{{\bf{W}}_{{\rm{\!BB,}}t}}\!{\bf{W}}_{{\rm{\!BB,}}t}^H\!{\bf{W}}_{{\rm{\!RF\!,}}t}^H{{\bf{A}}_{{\rm{R,G}}}}}.\!
	\end{align}

	In (\ref{Blk_opt_form_a}), for the optimization of digital combiner ${{\bf{W}}_{{\rm{BB,}}q}}$ within a single block, the overall block-diagonal constraint imposed on ${{\bf{W}}_{{\rm{BB}}}}$ can be ignored, then it turns to be an unconstrained optimization problem.
	However, the design of analog combiner with low-resolution PSs is still under non-convex constant modulus and discrete phase constraints, which cannot be solved optimally.
	Therefore, we optimize the analog and digital combiner separately, and alternating optimization is iteratively carried out between them.
	For $q = 1,2, ... ,N_{\rm{r}}^{{\rm{Block }}}$, we quantize ${{\bf{W}}_{{\rm{RF,}}q}}$ (corresponding to the $q$-th submatrix of $\bf{W}_\text{RF}$) rather than the whole $\bf{W}_\text{RF}$ in each iteration, which reliefs the performance loss effectively.
	Besides, thanks to the block-wise form in (\ref{block-wise_initial_obj}), the inevitable gap caused by phase quantization within a block can be compensated to some extent by the optimizations of subsequent blocks, thereby an improvement of the performance can be achieved.

	\subsection{Digital Combiner Design}\label{S5.1}

	First, we consider the design of ${{{\bf{W}}_{{\rm{BB,}}q}}}$ with fixed ${{{\bf{W}}_{{\rm{RF,}}q}}}$, which can be expressed as
	\begin{align} \label{Blk_BB_opt}
		\mathop {\text{min}}\limits_{{{\bf{W}}_{{\text{\!BB}},q}}}
		{\left\| {{\bf{A}}_{{\rm{R,G}}}^H{{\bf{W}}_{{\rm{\!RF\!,}}q}}{{\bf{W}}_{{\rm{\!BB,}}q}}{\bf{W}}_{{\rm{\!BB,}}q}^H{\bf{W}}_{{\rm{\!RF\!,}}q}^H{{\bf{A}}_{{\rm{R,G}}}} \!-\! {{\bf{E}}_q}} \right\|_F ^2}.\!
	\end{align}	
	The optimization problem in (\ref{Blk_BB_opt}) can also be solved according to (\ref{OptBB_Coh_opt_tot}-\ref{EigenDecom}).
	However, (\ref{EigenDecom}) will cause the performance loss to some extent when $N_\text{s} < N^\text{RF}$ and the difference among the eigenvalues of ${{\bf{X}}_{{\text{BB,}}q}}$ is relatively large.
	It is noted that the block-diagonal constraint imposed on ${\bf{W}}_{\text{BB}}$ is removed for the low-resolution PSs scenario by adopting the block-wise form. 
	Therefore, here we adopt the gradient descent algorithm to solve (\ref{EigenDecom})\cite{AdaptiveGD_SciDirectSP_2012}, which can achieve a better performance when $N_\text{s} < N^\text{RF}$.
	We define the objective function $\mathcal{J} $ as
	\begin{align} \label{GD_Obj_func}
		\mathcal{J}  \buildrel \Delta \over =  \left\| {{\bf{A}}_{\rm{E}}^H{{\bf{W}}_{{\rm{BB,}}q}}{\bf{W}}_{{\rm{BB,}}q}^H{{\bf{A}}_{\rm{E}}} - {{\bf{E}}_q}} \right\|_F^2,
	\end{align}	
	where ${{\bf{A}}_{\rm{E}}} = {\bf{W}}_{{\rm{RF,}}q}^H{{\bf{A}}_{{\rm{R,G}}}}$ denotes the equivalent array response matrix at UE.
	Then, we compute the gradient of $\mathcal{J}$ with respect to ${{{\bf{W}}_{{\rm{BB,}}q}}}$ as
	\begin{align} \label{}
		\nonumber \frac{{\partial {\cal J}}}{{\partial {{\bf{W}}_{{\rm{BB}},q}}}} \!= & \frac{\partial }{{\partial {{\bf{W}}_{{\rm{BB}},q}}}}{\mathop{\rm Tr}\nolimits} \left\{ {{{\left( {{\bf{A}}_{\rm{E}}^H{{\bf{W}}_{{\rm{BB}},q}}{\bf{W}}_{{\rm{BB}},q}^H{{\bf{A}}_{\rm{E}}} - {{\bf{E}}_q}} \right)}^H}} \right. \\
		\nonumber & \times \left. {\left( {{\bf{A}}_{\rm{E}}^H{{\bf{W}}_{{\rm{BB}},q}}{\bf{W}}_{{\rm{BB}},q}^H{{\bf{A}}_{\rm{E}}} - {{\bf{E}}_q}} \right)} \right\} \\
		\!= & \; 4{{\bf{A}}_{\rm{E\!}}}\!\left(\! {{\bf{A}}_{\rm{E\!}}^H{{\bf{W}}_{{\rm{\!BB,}}q}}\!{\bf{W}}_{{\rm{\!BB,}}q}^H{{\bf{A}}_{\rm{E\!}}} \!-\! {{\bf{E}}_q}} \!\right)\!\!{\bf{A}}_{\rm{E}}^H{{\bf{W}}_{{\rm{\!BB,}}q}}.\!
	\end{align}		
	Thus, the updating equation can be given by
	\begin{align} \label{W_BB_k+1_iter}
		\nonumber {{\bf{W}}_{{\rm{BB,}}q,k + 1}} =\; & {{\bf{W}}_{{\rm{BB,}}q,k}} - {\eta _{k}}{{\bf{A}}_{\rm{E}}}\left( {\bf{A}}_{\rm{E}}^H{{\bf{W}}_{{\rm{BB,}}q,k}}{\bf{W}}_{{\rm{BB,}}q,k}^H{{\bf{A}}_{\rm{E}}} \right. \\
		& \left. - {{\bf{E}}_q} \right){\bf{A}}_{\rm{E}}^H{{\bf{W}}_{{\rm{BB,}}q,k}},
	\end{align}		
	where $k$ is the iteration index and ${\eta _{k}}$ denotes the stepsize in the $k$-th iteration.
	In order to accelerate the convergence speed and achieve a better performance, the steepest descent method is used which can be described as follows \cite{AdaptiveGD_SciDirectSP_2012}:
	\begin{align} \label{}
		{\eta _{k + 1}} = {\eta _k} - \gamma \frac{{\partial {\cal J}\left( {{{\bf{W}}_{{\rm{BB,}}q,k + 1}}} \right)}}{{\partial {\eta _k}}},
	\end{align}
	where ${{\cal J}\left( {{{\bf{W}}_{{\rm{BB,}}q,k + 1}}} \right)}$ denotes the value of objective function by substituting (\ref{W_BB_k+1_iter}) in the $(k+1)$-th iteration, and $\gamma$ is a constant \cite{AdaptiveGD_SciDirectSP_2012}. 
	By expanding the Frobenius norm, we calculate that
	\begin{align} \label{Ada_stepsize}
		\nonumber & \; \frac{\partial}{{\partial {\eta _k}}}{\cal J}\left( {{{\bf{W}}_{{\rm{BB,}}q,k + 1}}} \right) \\
		\nonumber = & \;\frac{\partial }{{\partial {\eta _k}}}\left\| {{\bf{A}}_{\rm{E}}^H{{\bf{W}}_{{\rm{BB,}}q,k + 1}}{\bf{W}}_{{\rm{BB,}}q,k + 1}^H{{\bf{A}}_{\rm{E}}} \!-\! {{\bf{E}}_q}} \right\|_F^2 \\
		\nonumber = & \; 4\eta _k^3{\rm{Tr}}\left\{ {{{\bf{\Gamma }}_2}{{\bf{\Gamma }}_2}} \right\} - 3\eta _k^2{\rm{Tr}}\left\{ {{{\bf{\Gamma }}_2}{\bf{A}}_{\rm{E}}^H{{\bf{\Gamma }}_3}{{\bf{A}}_{\rm{E}}} + {\bf{A}}_{\rm{E}}^H{{\bf{\Gamma }}_3}{{\bf{A}}_{\rm{E}}}{{\bf{\Gamma }}_2}} \right\} \\
		\nonumber & - 2{\eta _k}{\mathop{\rm Tr}\nolimits} \left\{ {{\bf{\Gamma }}_4^H{{\bf{\Gamma }}_4} + {{\bf{\Gamma }}_4}{{\bf{\Gamma }}_4}{\rm{ + }}{{\bf{\Gamma }}_2}{\bf{A}}_{\rm{E}}^H{{\bf{W}}_{{\rm{BB}},q,k}}{\bf{W}}_{{\rm{BB}},q,k}^H{{\bf{A}}_{\rm{E}}}} \right. \\
		\nonumber & \left. { + {\bf{A}}_{\rm{E}}^H{{\bf{\Gamma }}_3}{{\bf{A}}_{\rm{E}}}{\bf{A}}_{\rm{E}}^H{{\bf{W}}_{{\rm{BB}},q,k}}{\bf{W}}_{{\rm{BB}},q,k}^H{{\bf{\Gamma }}_1}{{\bf{A}}_{\rm{E}}} - {{\bf{\Gamma }}_2}{{\bf{E}}_q}} \right\} \\
		& +\! {\rm{\!Tr\!}}\left\{{{{\bf{\!A}}_{\rm{\!E}}}{{\bf{\Gamma }}_{\!3}}{\bf{A}}_{\rm{\!E}}^H{{\bf{\!E}}_q} \!-\! {{\bf{\!A}}_{\rm{\!E}}}{{\bf{\Gamma }}_{\!3}}{{\bf{A}}_{\rm{\!E}}}{\bf{A}}_{\rm{\!E}}^H{{\bf{\!W}}_{{\rm{\!\!BB\!}},q\!,k}}{\bf{\!W}}_{{\rm{\!\!BB\!}},q\!,k}^H{{\bf{A}}_{\rm{\!E}}}} \right\}\!,\!
	\end{align}
	where ${{\bf{\Gamma }}_1}$, ${{\bf{\Gamma }}_2}$ and ${{\bf{\Gamma }}_3}$ are Hermittian matrices except ${{\bf{\Gamma }}_4}$, and we have
	\begin{align} \label{}
		\nonumber & {{\bf{\Gamma }}_1} \buildrel \Delta \over = {{\bf{A}}_{\rm{E}}}\left( {{\bf{A}}_{\rm{E}}^H{{\bf{W}}_{{\rm{BB,}}q,k}}{\bf{W}}_{{\rm{BB,}}q,k}^H{{\bf{A}}_{\rm{E}}} - {{\bf{E}}_q}} \right){\bf{A}}_{\rm{E}}^H, \\
		\nonumber & {{\bf{\Gamma }}_2} \buildrel \Delta \over = {\bf{A}}_{\rm{E}}^H{{\bf{\Gamma }}_1}{{\bf{W}}_{{\rm{BB,}}q,k}}{\bf{W}}_{{\rm{BB,}}q,k}^H{{\bf{\Gamma }}_1}{{\bf{A}}_{\rm{E}}}, \\
		\nonumber &  {{\bf{\Gamma }}_3} \buildrel \Delta \over = {{\bf{\Gamma }}_1}{{\bf{W}}_{{\rm{BB,}}q,k}}{\bf{W}}_{{\rm{BB,}}q,k}^H + {{\bf{W}}_{{\rm{BB,}}q,k}}{\bf{W}}_{{\rm{BB,}}q,k}^H{{\bf{\Gamma }}_1}, \\
		& {{\bf{\Gamma }}_4} \buildrel \Delta \over = {\bf{A}}_{\rm{E}}^H{{\bf{\Gamma }}_1}{{\bf{W}}_{{\rm{BB,}}q,k}}{\bf{W}}_{{\rm{BB,}}q,k}^H{{\bf{A}}_{\rm{E}}}.
	\end{align}	
	Therefore, we can update ${{{\bf{W}}_{{\rm{BB,}}q}}}$ with an adaptive stepsize until the $\mathcal{J}$ in (\ref{GD_Obj_func}) no longer decreases.

	\subsection{Analog Combiner Design}\label{S5.2}

	The design of analog combiner ${{{\bf{W}}_{{\rm{RF,}}q}}}$ with low-resolution PSs in the $q$-th block is considered in this subsection when ${{{\bf{W}}_{{\rm{BB,}}q}}}$ is fixed, which can be expressed as
	\begin{subequations} \label{Blk_RF_opt}
		\begin{align}
			\mathop {\text{min}}\limits_{{{\bf{W}}_{{\text{RF\!}},q}}} & 
			{\left\| {{\bf{A}}_{{\rm{R,G}}}^H{{\bf{W}}_{{\rm{\!RF,}}q}}{{\bf{W}}_{{\rm{\!BB,}}q}}{\bf{W}}_{{\rm{\!BB,}}q}^H{\bf{W}}_{{\rm{\!RF,}}q}^H{{\bf{A}}_{{\rm{R,G}}}} \!-\! {{\bf{E}}_q}} \right\|_F ^2} \label{Blk_RF_opt_form_a} \\
			\text{s.t. }& \; {{\mathbf{W}}_{{\text{\!RF}},q}}(i,j)\in \mathcal{W},\forall i,j \label{Blk_RF_opt_cons_b}.
		\end{align}
	\end{subequations}	

	Traditional optimization methods cannot be applied to aforementioned discrete non-convex optimization problem directly, and it is impossible to obtain the global optimal solution straightforward.
	Consequently, we propose to temporarily ignore the discrete-phase constraints caused by the resolution of PSs, and obtain the approximate optimal analog combiner with continuous phase by alternately optimizing between analog and digital combiner after a sufficient number of iterations.
	Then, the approximate optimal analog combiner with discrete-phase can be obtained by quantizing the phase of above continuous-phase solution.
	
	Considering the similarity between (\ref{OptRF_Coh_opt_tot}) and (\ref{Blk_RF_opt}) except for the discrete-phase constraints in (\ref{Blk_RF_opt_cons_b}), (\ref{Blk_RF_opt}) can also be solved by Algorithm 1 with minor changes.
	In addition to the changes of matrix/vector dimension caused by block-wise formulation, the cost function of (\ref{Blk_RF_opt_form_a}) becomes
	\begin{align}\label{Blk_cost_func}
		f({{\bf{x}}_q}) \!\!\buildrel \Delta \over =\!\! \left\| {\bf{\!A}}_{{\rm{\!R,\!G}}}^H{\rm{invec\!}}\left( {{{\bf{x}}_q}} \right){{\bf{\!W}}_{{\rm{\!\!BB,}}q}}{\bf{\!W}}_{{\rm{\!\!BB,}}q}^H {{\left( {{\rm{invec\!}}\left( {{{\bf{x}}_q}} \right)} \right)}^H}{{\bf{\!\!A}}_{{\rm{\!R,\!G}}}} \!- {{\bf{\!E}}_q} \right\|_{\!F}^2\!,\!
	\end{align}		
	where ${\rm{invec}}\left( {{{\bf{x}}_q}} \right) \in {\mathbb{C}^{{N_{\rm{r}}} \times {N^{{\rm{RF}}}}}}$ corresponds to the analog combiner ${{\bf{W}}_{{\rm{RF,}}q}}$, then its Euclidean gradient changes to be
	\begin{align}\label{Blk_Euc_Gra}
		\nonumber \!\nabla \!f\!\left({{{\bf{x}}_q}} \right) \!= & 4{\mathop{\rm vec\!}\nolimits} \left[ {{{\bf{\!A}}_{{\rm{\!R}},{\rm{G\!}}}}\!\left( {{\bf{\!\!A}}_{{\rm{\!R}},{\rm{G}}}^H{\mathop{\rm invec\!}\nolimits} \left( {{{\bf{x}}_q}} \right){{\bf{\!\!W}}_{{\rm{\!BB\!}},q}}{\bf{\!W}}_{{\rm{\!BB\!}},q}^H{{\left( {{\mathop{\rm invec\!}\nolimits} \left( {{{\bf{x}}_q}} \right)} \right)}^{\!H}}} \right.} \right. \\
		& \!\!\! \left.{\left.{\times \;{{\bf{\!A}}_{{\rm{R}},{\rm{G}}}} \!-\! {{\bf{E}}_q}} \right){\bf{\!A}}_{{\rm{R}},{\rm{G}}}^H{\rm{ invec\! }}\left( {{{\bf{x}}_q}} \right){{\bf{\!W}}_{{\rm{\!BB}},q}}{\bf{\!W}}_{{\rm{\!BB}},q}^H} \right].
	\end{align}
	Then we can similarly proceed it according to Algorithm 1. 
	After obtaining a stable continuous-phase analog combiner, the final discrete-phase ${{{\bf{W}}_{{\rm{RF,}}q}}}$ can be obtained by the quantization function ${\cal Q}( \cdot )$ which quantizes a complex unit-norm variable to the nearest point in the set $\mathcal{W}$.

	\subsection{Hybrid Combiner Design}\label{S5.3}

	In the above two subsections, for $q = 1,2, ... ,N_{\rm{r}}^{{\rm{Block }}}$, we solve the optimization problems of ${{{\bf{W}}_{{\rm{BB,}}q}}}$ and ${{{\bf{W}}_{{\rm{RF,}}q}}}$ when the other is fixed.
	Furthermore, a hybrid combiner design algorithm with low-resolution PSs is proposed as follows, which proceeds alternating optimization between analog/digital sensing matrices and loops among blocks.
	\begin{algorithm}[th!]
		\renewcommand{\algorithmicrequire}{\textbf{Input:}}
		\renewcommand\algorithmicensure {\textbf{Output:} }
		\caption{Proposed Block-wise Alternating Hybrid Analog-Digital Combiner Design Algorithm with Low-Resolution PSs}
		\label{}
		\begin{algorithmic}[1]
			\STATE\textbf{Input:} ${{{\bf{A}}_{{\rm{R,G}}}}}$;
			\STATE \textbf{Initialization:}
			Set $\bf{W}_\text{BB}$ with random initial value $\bf{W}_\text{BB,0}$, construct $\bf{W}_\text{RF}$ with random phases $\bf{W}_\text{RF,0}$ chosen from $\mathcal{W}$, $q=1$; \\
			\REPEAT
			\STATE Calculate ${\bf{E}}_q$ by (\ref{E_q}) and set $k=0$;
			\STATE Obtain ${{\bf{W}}_{{\rm{BB,}}q,0}}$ and ${{\bf{W}}_{{\rm{RF,}}q,0}}$ (the $q$-th submatrices of ${{\bf{W}}_{{\rm{BB}}}}$ and ${{\bf{W}}_{{\rm{RF}}}}$, respectively);
			\REPEAT	\label{blk_opt_begin}		
			\STATE Optimize ${{\bf{\!W}}_{{\text{\!\!BB,}}q,k+\!1}}$ with fixed ${{\bf{\!W}}_{{\text{\!\!RF,}}q,k}}$ by solving \!(\ref{Blk_BB_opt});\!
			\STATE Fix ${{\bf{W}}_{{\text{BB,}}q,k+1}}$, obtain ${{\bf{W}}_{{\text{RF,}}q,k+1}}$ according to (\ref{Blk_RF_opt}-\ref{Blk_Euc_Gra}) and Algorithm 1;
			\STATE $k \leftarrow k + 1$ ;
			\UNTIL a stopping criterion triggers. \label{blk_opt_end}	
			\STATE ${{\bf{W}}_{{\rm{RF,}}q}}={\cal Q}({{\bf{W}}_{{\rm{RF,}}q,{\rm{k - 1}}}})$; \label{Qua_step}
			\STATE Obtain ${{\bf{W}}_{{\text{BB,}}q}}$ with fixed ${{\bf{W}}_{{\text{RF,}q}}}$ by solving (\ref{Blk_BB_opt}); \label{Blk_BB_opt_Qua}
			\IF {the value of (\ref{Blk_opt_form_a}) decreases after the update in the $q$-th block} \label{judge_step_begin}
			\STATE Replace the corresponding part of ${{\bf{W}}_{{\text{BB}}}}$ and ${{\bf{W}}_{{\text{RF}}}}$ with ${{\bf{W}}_{{\text{BB,}}q}}$ and ${{\bf{W}}_{{\text{RF,}q}}}$, respectively;
			\ENDIF \label{judge_step_end}
			\STATE $q \leftarrow \bmod (q,N_{\rm{r}}^{{\rm{Block }}}) + 1$; \label{Blk_loop}
			\UNTIL a stopping criterion triggers. \label{Blk_loop_end}			
			\STATE Obtain the final hybrid combiner ${{\bf{W}}_{{\rm{RF}}}}$ and ${{\bf{W}}_{{\rm{BB}}}} = \frac{{\sqrt {{T_\text{r}}} {{\bf{W}}_{{\rm{BB}}}}}}{{{{\left\| {{{\bf{W}}_{{\rm{RF}}}}{{\bf{W}}_{{\rm{BB}}}}} \right\|}_F}}}$; \label{Blk_Power_cons}
			\STATE\textbf{Output:} ${{\bf{W}}_{{\rm{RF}}}}$, ${{\bf{W}}_{{\rm{BB}}}}$ \\
		\end{algorithmic}
	\end{algorithm}	

	The overall algorithm is to iterate between the design of ${{{\bf{W}}_{{\rm{BB,}}q}}}$ and ${{{\bf{W}}_{{\rm{RF,}}q}}}$ among $N_{\rm{r}}^{{\rm{Block }}}$ blocks.
	The design of the analog sensing submatrix is first with continuous phase (Step \ref{blk_opt_begin}-\ref{blk_opt_end}), where the iteration in Step \ref{blk_opt_end} stops when the objective function (\ref{Blk_opt_form_a}) no longer decreases.
	Then the analog sensing submatrix is quantized to a discrete-phase form determined by the resolution of PSs (Step \ref{Qua_step}), and the digital sensing submatrix design is adjusted accordingly (Step \ref{Blk_BB_opt_Qua}).

	Note that the quantization step inevitably causes the performance loss to some extent, so the achievable performance after optimizing the  hybrid sensing matrix in the $q$-th block may not necessarily be improved compared with that before updating.
	Therefore, for the purpose of avoiding performance degradation caused by quantization, a judgement step is added in Step \ref{judge_step_begin}-\ref{judge_step_end} to ensure the monotonicity of the value of objective function during iterations.

	Besides, since the value of ${\bf{E}}_q$ in (\ref{Blk_opt_form_a}) is related to all other blocks in the optimization process of the $q$-th block, 
	we take the modulus and remainder operations of $q$ to $N_{\rm{r}}^{{\rm{Block }}}$ in Step \ref{Blk_loop} to loop among $N_{\rm{r}}^{{\rm{Block }}}$ blocks.
	In Step \ref{Blk_loop_end}, the iteration stops when $N_{\rm{r}}^{{\rm{Block }}}$ consecutive blocks have not been effectively updated after judgments. After scaling ${{{\bf{W}}_{{\rm{BB}}}}}$ to meet the power constraint (Step \ref{Blk_Power_cons}), the final hybrid combiner with low-resolution PSs can be obtained.

	The complexity analysis of Algorithm 3 for low-resolution PSs scenario is given as follows.
	The computational complexity is mainly composed of two parts, namely the digital combiner design and the analog combiner design.
	The updating of digital sub-combiner ${\bf{W}}_{\text{BB},q}$ in (\ref{Blk_BB_opt}) needs about $\mathcal{O}\left(N_\text{RF} G_\text{r}^2\right)$ computations for each iteration. 
	Thus, the digital sub-combiner can be obtained with computational complexity $\mathcal{O}\left(N_\text{iter,2}^\text{d} N_\text{RF} G_\text{r}^2\right)$, where $N_\text{iter,2}^\text{d}$ denotes the number of iterations for the gradient descent algorithm (corresponding to Step 7 in Algorithm 3). 
	Similar to the corresponding part of Algorithm 2, the computational complexity of analog sub-combiner design is $\mathcal{O} \left( N_\text{iter,2}^\text{a} N_\text{r} G_\text{r}^2 \right)$, where $N_\text{iter,2}^\text{a}$ denotes the number of iterations (corresponding to Step 8 in Algorithm 3).
	Therefore, the overall computational complexity of Algorithm 3 is about $\mathcal{O}\left(N_\text{iter}^\text{o} N_\text{iter}^\text{i} (N_\text{iter,2}^\text{d} N_\text{RF} G_\text{r}^2 + N_\text{iter,2}^\text{a} N_\text{r} G_\text{r}^2) \right)$, where $N_\text{iter}^\text{o}$ and $N_\text{iter}^\text{i}$ denote the number of outer and inner iterations in Algorithm 3, respectively.

    \section{Simulation Results}\label{S6}

	In this section, we provide simulation results to evaluate the performance of proposed algorithms. 
	The simulation scenarios and relevant parameters are shown as follows:

	\textit{1) System Parameters:} A point-to-point mmWave MIMO system with hybrid analog-digital structures is considered in this paper, where we consider that ULAs are equipped at both UE and BS with ${N_{\text{r}}}=32$ and ${N_{\text{t}}}=64$. 
	We assume that the number of RF chains at UE and BS are equal, and the RF chains number of $N^\text{RF} = 4$ is simulated. 
	Besides, we set the number of the space between antennas equal to $\lambda /2$.
	The PS resolutions of $B = 1,2,3$ bits are considered as low-resolution, contrasting to the infinite-resolution of $B = \infty $.

	\textit{2) Channel Parameters:} We adopt the channel model defined in (\ref{xishuxindaojuzhen}) with $L=4$ and the path gain ${\alpha _l} \sim {\cal C}{\cal N}\left( {0,\frac{1}{L}} \right)$. 
	The AoAs/AoDs are assumed to be non-uniformly distributed in $\left[ {0,\pi } \right]$ as defined in (\ref{xishuxindaojuzhen}) \cite{CE_OMP_Korea_TCOM_2016}.
	To make the mmWave channel sparse enough, the angular grid resolutions need to be set higher than the number of antennas at UE and BS with ${G_{\text{r}}}=36$ and ${G_{\text{t}}}=72$, respectively. 
	The system is assumed to work in a scenario with 28 GHz carrier frequency and 100 MHz bandwidth \cite{CE_Ahmed_JSTSP_2014}.

	\textit{3) Relevant Definitions:} We adopt the NMSE to evaluate the accuracy of channel estimation, which can be defined as $10{\text{lo}}{{\text{g}}_{10}}\left( \mathbb{E}{\left[ {\left\| {{\bf{H}} - {\bf{\hat H}}} \right\|_F^2/\left\| {\bf{H}} \right\|_F^2} \right]} \right)$, where ${\bf{H}}$ and ${{\bf{\hat H}}}$ denote the generated channel matrix and estimated channel matrix, respectively, 
	and ${{\bf{\hat H}}}$ is obtained by using the OMP based channel estimator proposed in \cite{CE_OMP_Korea_TCOM_2016}.
	Inspired by \cite{CE_OMP_Korea_TCOM_2016}, two types of SNRs are considered in this paper, which can be defined as pilot-to-noise ratio (PNR) equal to $10{\log _{10}}\left( {{P_{{\text{ce}}}}/\sigma _n^2} \right)$ and data-to-noise ratio (DNR) equal to $10{\log _{10}}\left( {{P_{{\text{hp}}}}/\sigma _n^2} \right)$, where ${{P_{{\text{ce}}}}}$ and ${{P_{{\text{hp}}}}}$ represent the pilot power for channel estimation and the data power for hybrid precoding, respectively.
	For the purpose of evaluating the performance of channel estimation, we further proceed hybrid precoding using the algorithm proposed in \cite{LowPSPrecoding_YuWei_JSTSP_2015} according to the estimated channel matrices, and the achievable spectral frequency can be given by $R = {\log _2}\left| {{{\bf{I}}_{{N_{\text{s}}}}} + \frac{{{P_{{\text{hp}}}}}}{{{N_{\text{s}}}}}{{\left( {{{{\bf{\tilde W}}}^H}{\bf{\tilde W}}} \right)}^{ - 1}}{{{\bf{\tilde W}}}^H}{\bf{H\tilde F}}{{{\bf{\tilde F}}}^H}{{\bf{H}}^H}{\bf{\tilde W}}} \right|$, where ${{\bf{\tilde F}}}$ and ${{\bf{\tilde W}}}$ denote the hybrid precoder and combiner designed with specific-resolution PSs using the estimated channel matrix ${{\bf{\hat H}}}$.

	Then, the proposed algorithms will be compared with some existing schemes in the first subsection, and in the second subsection, the simulation results will be provided to verify the convergence and the universality of proposed algorithms under different system parameters.
	The results in this simulation are obtained over 500 channel realizations. 
	Besides, in Algorithm 2 and Algorithm 3, the initial analog sensing matrix is generated randomly with continuous/discrete phase under the constant modulus constraints.
	Under the block-diagonal constraint, the initial digital submatrix within a single block is a complex random matrix.
	The distribution of initial random hybrid sensing matrix will not make a significant impact on the final performance.

	\subsection{Performance Comparison}\label{S6.1}

	First of all, for the infinite-resolution PSs scenario, we compare the proposed Algorithm 2 with the methods in \cite{CE_Model_Korea_GCC_2014,CE_OMP_Korea_TCOM_2016}.
	Both of the two literatures provide designs of analog-digital sensing matrices, and good performances of channel estimation can be achieved.
	However, the system models in \cite{CE_Model_Korea_GCC_2014,CE_OMP_Korea_TCOM_2016} are with some specific constraints\footnote{Specifically, taking the receiving side as an example, the digital sub-combiner ${\bf{W}}_{\text{BB},q}$ in \cite{CE_Model_Korea_GCC_2014} must be a square matrix (i.e. $N_\text{s} = N^\text{RF}$), and so is the analog combiner ${\bf{W}}_{\text{RF}}$ in \cite{CE_OMP_Korea_TCOM_2016} (i.e. $M_\text{r} =N_\text{r}$, $T_\text{r} = \frac{N_\text{r} N_\text{s}}{N^\text{RF}}$). In order to overcome the constraints imposed on ${\bf{W}}_{\text{BB},q}$ and ${\bf{W}}_{\text{RF}}$, we consider a more generalized model with $N_\text{s} \leq N^\text{RF}$ in this paper. Besides, the choice of training beam number is flexible, which  is unique for fixed $N_\text{r}$, $N_\text{s}$, and $N^\text{RF}$ in \cite{CE_OMP_Korea_TCOM_2016} (i.e. $T_\text{r} = \frac{N_\text{r} N_\text{s}}{N^\text{RF}}$).}. 
	Therefore, it is necessary for us to consider a more generalized model  in this paper.

	\begin{figure}[t]
		\center{\includegraphics[width=0.5\textwidth]{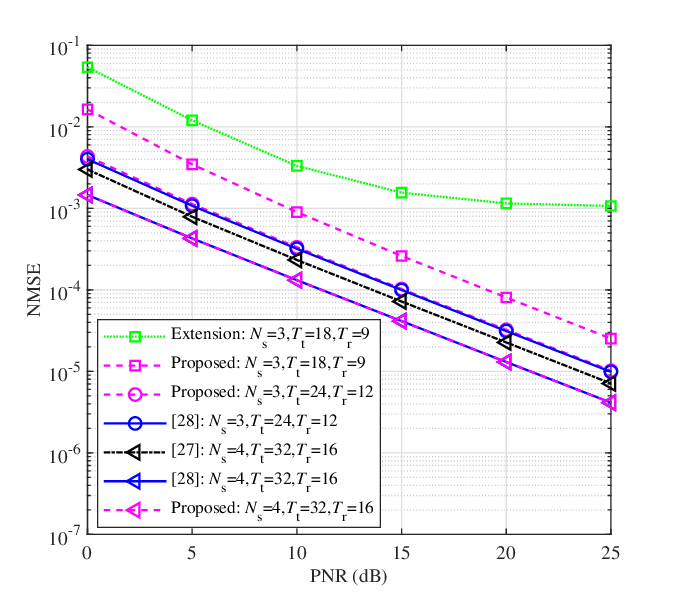}}
			\caption{NMSE performance against PNR using several types of training beam patterns when ${N_{\text{t}}} = 32$, ${N_{\text{r}}} = 16$, $N^\text{RF} = 4$, and $B = \infty$.}
		\label{fig3}
	\end{figure}
	Fig. \ref{fig3} shows the NMSEs of channel estimation against PNR. 
	In order to form an effective comparison in a generalized scenario (i.e. we assume $N_\text{s} \leq N^\text{RF}$, and there is no need for $T_\text{r} = \frac{N_\text{r} N_\text{s}}{N^\text{RF}}$), we extend the methods proposed in existing literatures (denoted as `Extension'), where the analog  and digital sensing matrices are designed according to literature \cite{CE_Model_Korea_GCC_2014} and \cite{CE_OMP_Korea_TCOM_2016}, respectively.
	Partial-training and full-training mean the number of training beams is less than or equal to that of antennas \cite{CE_OMP_Korea_TCOM_2016}, respectively.
	As shown in Fig. \ref{fig3}, the achievable NMSE performances of proposed Algorithm 2 and the method in \cite{CE_OMP_Korea_TCOM_2016} are relatively optimal when $N_\text{s} = N^\text{RF}$ with full-training ($N_\text{s}=4$, $T_\text{t}=32$, $T_\text{r}=16$).
	Besides, both of the two methods above achieve better performance than that of the method in \cite{CE_Model_Korea_GCC_2014}.
	Although the NMSE performances of proposed Algorithm 2 and the method in \cite{CE_OMP_Korea_TCOM_2016} are almost the same when $N_\text{s} < N^\text{RF}$ ($N_\text{s}=3$, $T_\text{t}=24$, $T_\text{r}=12$), 
	the NMSE performance of the proposed method is much better than that of the extension (denoted as `Extension') of literatures \cite{CE_Model_Korea_GCC_2014,CE_OMP_Korea_TCOM_2016} in a more generalized scenario with partial-training ($N_\text{s}=3$, $T_\text{t}=18$, $T_\text{r}=9$).

	In the following, we consider six types of training beam patterns in a generalized scenario, which can be described as:

	$i$) For the mmWave communication system with full-digital structure, we use the eigenvalue decomposition based method proposed in \cite{SensingOpti_Block_Israel_TSP_2011} to construct the sensing matrix for channel estimation.
	Since no PSs is used in the full-digital structure, the design of sensing matrix will not be constrained by constant modulus and discrete phase with the highest degree of freedom;

	$ii$) The hybrid sensing matrix is designed using the proposed Algorithm 2 with infinite-resolution PSs;

	$iii$) To form an effective comparison with the proposed Algorithm 2, the alternating minimization (AltMin) algorithm proposed in \cite{ManifoldOpti_AltMin_JSTSP_2016} is also simulated under the scenario with infinite-resolution PSs, in which the alternate iteration is performed alternately between the analog and digital sensing matrix design.
	The full-digital precoder/combiner of type $i$) is used as the approaching target in the AltMin algorithm;

	$iv$) The hybrid sensing matrix is designed using the proposed block-wise alternating Algorithm 3 with low-resolution PSs;

	$v$) Similarly, here we also provide an intuitive way to design the low-resolution hybrid sensing matrix based on existing methods as an benchmark, where the low-resolution analog sensing matrix is obtained by quantizing the infinite-resolution sensing matrix of type $iii$) directly.
	Then the block-diagonal digital sensing matrix is designed block-by-block, where the digital submatrix can be obtained by multiplying the pseudo-inverse of the resultant analog part by the hybrid submatrix of type $iii$) within a single block.
	In the simulation figures, this method is denoted as alternating minimization with direct quantization (AltMin-DQ);

	$vi$) The random scheme (denoted as `Random') in \cite{LowPSCE_BeiDa_VTC_2017} is used to construct the low-resolution analog sensing matrix, where its elements are generated randomly and uniformly with quantized phases, and the design of digital sensing matrices is exactly the same as that of type $v$).

	In addition, the scheme of using DFT matrices to construct analog sensing matrices in \cite{CE_Model_Korea_GCC_2014,CE_OMP_Korea_TCOM_2016} is not applicable to the scenario where the number of training beams is less than that of antennas, so it is not added in the simulation.
	\begin{figure}[t]
		\centering
		\subfigure[]{\includegraphics[width=0.495\textwidth]{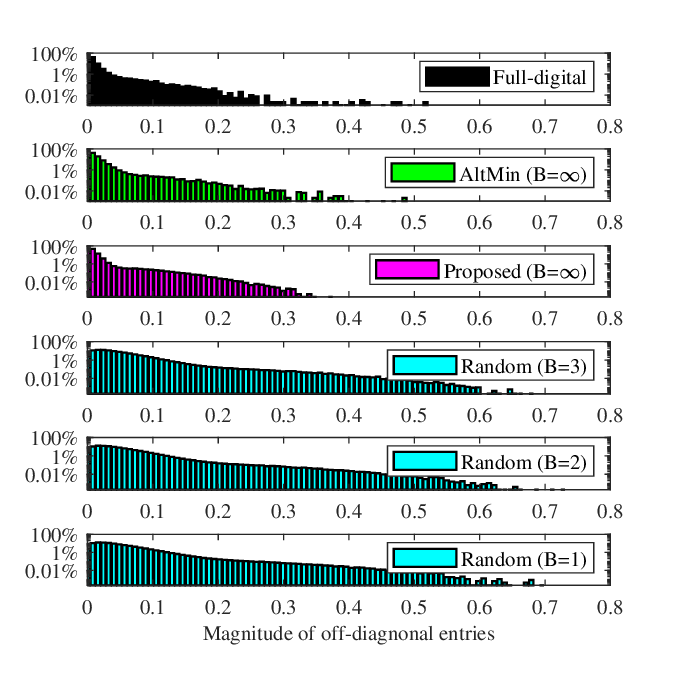}}
		\subfigure[]{\includegraphics[width=0.495\textwidth]{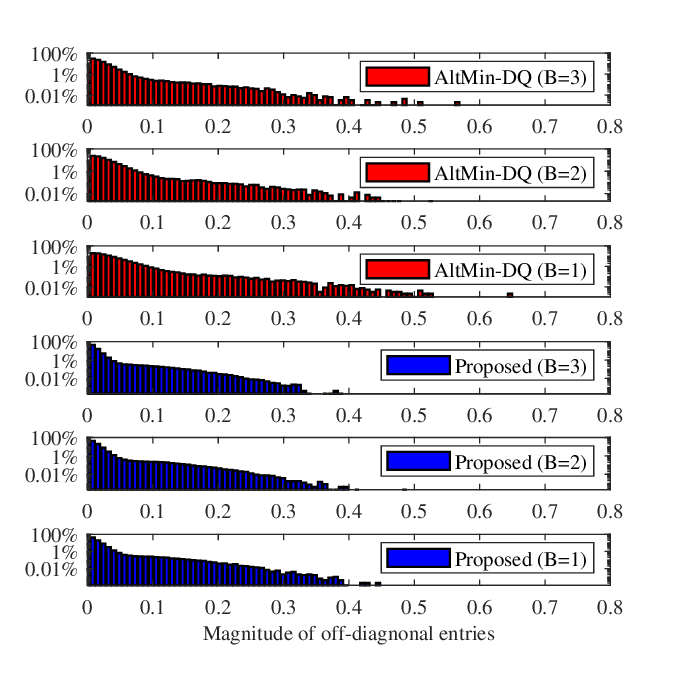}}
		\caption{Histograms of the magnitudes of off-diagonal entries of normalized Gram sensing matrices using six types of training beam patterns when ${N_{\text{t}}} = 64$, ${N_{\text{r}}} = 32$, $N_\text{s}=N^\text{RF} = 4$, $T_\text{t} = 48$, $T_\text{r} = 24$, ${G_{\text{t}}} = 72$, ${G_{\text{r}}} = 36$, and $B = 1,2,3,\infty$.}
		\label{fig4}
	\end{figure}
	\begin{figure}[t]
		\center{\includegraphics[width=0.5\textwidth]{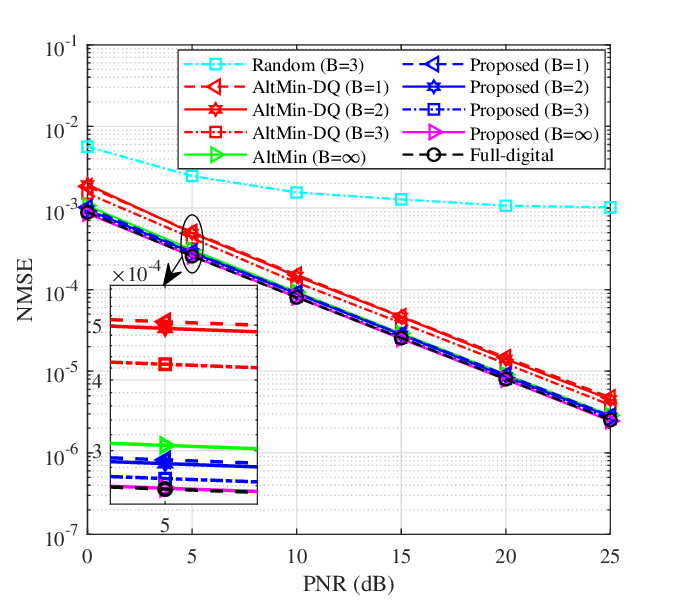}}
		\caption{NMSE performance against PNR using six types of training beam patterns when ${N_{\text{t}}} = 64$, ${N_{\text{r}}} = 32$, $N_\text{s}=N^\text{RF} = 4$, $T_\text{t} = 48$, $T_\text{r} = 24$, ${G_{\text{t}}} = 72$, ${G_{\text{r}}} = 36$, and $B = 1,2,3,\infty$.}
		\label{fig5}
	\end{figure}

	In order to show the column coherence distribution of the designed sensing matrix $\bf{Q}$ as defined in (\ref{Q_def}),
	Fig. \ref{fig4} shows histograms of the magnitudes of off-diagonal entries of normalized Gram matrices, where the normalized Gram of equivalent dictionary matrices are constructed using the aforementioned six schemes.
	The values of column coherence of equivalent dictionary matrix are between 0 and 1, therefore we aim for making their distributions closer to 0 (the left side on the horizontal axis) to achieve a better recovery performance of compressive sensing algorithms.
	It can be seen that although there is still a small gap compared with the full-digital scheme, the schemes using proposed algorithms (Algorithm 2 and Algorithm 3) can achieve better coherence distributions than that of AltMin algorithm with infinite-resolution PSs and AltMin-DQ algorithm with low-resolution PSs, respectively.
	Besides, the random scheme achieves the worst performance, and there is no obvious difference among $B = 1,2,3$, so we only consider the case of $B = 3$ in some of the following simulation figures.

	As shown in Fig. \ref{fig5}, the NMSEs (except random scheme) of channel estimation decrease exponentially with the increase of PNR, where the full-digital scheme can achieve the most accurate estimation because it is free of constant modulus and discrete phase constraints.
	Under the condition of infinite-resolution PSs, the proposed Algorithm 2 can achieve almost exact the same NMSE performance as that of full-digital scheme, and exceeds the NMSE performance of AltMin ($B = \infty$) algorithm significantly.
	Besides, compared with the AltMin-DQ algorithm and the random scheme, the proposed Algorithm 3 can achieve lower NMSEs with $B = 1,2,3$,
	and the NMSE performance shows obvious monotonicity against resolutions of PSs, 
	because there is a higher degree of freedom for the design of hybrid sensing matrix with higher-resolution PSs.
	\begin{figure}[t]
		\center{\includegraphics[width=0.5\textwidth]{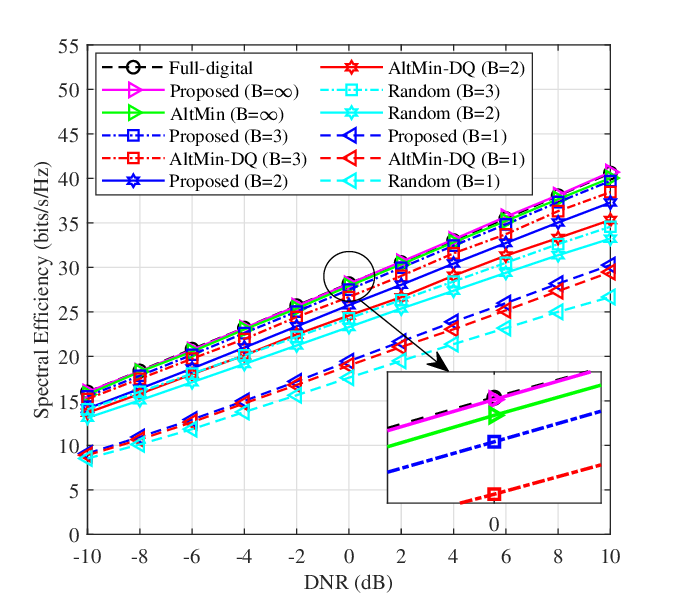}}
		\caption{Spectral frequency for precoding against DNR using the channel  matrices estimated by six types of training beam patterns when ${N_{\text{t}}} = 64$, ${N_{\text{r}}} = 32$, $N_\text{s}=N^\text{RF} = 4$, $T_\text{t} = 48$, $T_\text{r} = 24$, ${G_{\text{t}}} = 72$, ${G_{\text{r}}} = 36$, $\text{PNR = -10 dB}$ and $B = 1,2,3,\infty$.}
		\label{fig6}
	\end{figure}

	Fig. \ref{fig6} shows the achievable spectral frequency for precoding against DNR.
	It should be noted that the objective of hybrid precoding (maximizing the spectral efficiency) is totally different with that of channel estimation (MIP), so the corresponding hybrid precoding and combining  matrices need to be redesigned.
	For the full-digital case, optimal $N_\text{s}$-stream full-digital linear precoder and combiner is designed according to the singular value decomposition of the estimated channel matrix using the eigenvalue decomposition based method in \cite{LowPSPrecoding_YuWei_JSTSP_2015}, where the sensing matrices of type $i$) are used for channel estimation. 
	For the hybrid analog-digital case ($B = 1,2,3,\infty$), we use corresponding methods proposed in this paper to design the hybrid sensing matrix, then the estimated channel matrix is used to proceed with hybrid precoding, where the hybrid precoding and combining matrices are designed according to the methods proposed in \cite{LowPSPrecoding_YuWei_JSTSP_2015}.
	In Fig. \ref{fig6}, it can be seen that higher spectral efficiency can be obtained according to the channel estimated with well-designed sensing matrix. 
	Moreover, the performance of the proposed Algorithm 2 with $B=\infty$ is very close to the achievable spectral frequency of optimal full-digital precoding scheme, which is nearly optimal.
	In the case of low-resolution PSs, the achievable spectral frequency using the proposed Algorithm 3 is higher than that using AltMin-DQ and random scheme with $B=1,2,3$, respectively.

    \subsection{Analysis for The Proposed Algorithms}\label{S6.2}
	\begin{figure}[t]
		\center{\includegraphics[width=0.5\textwidth]{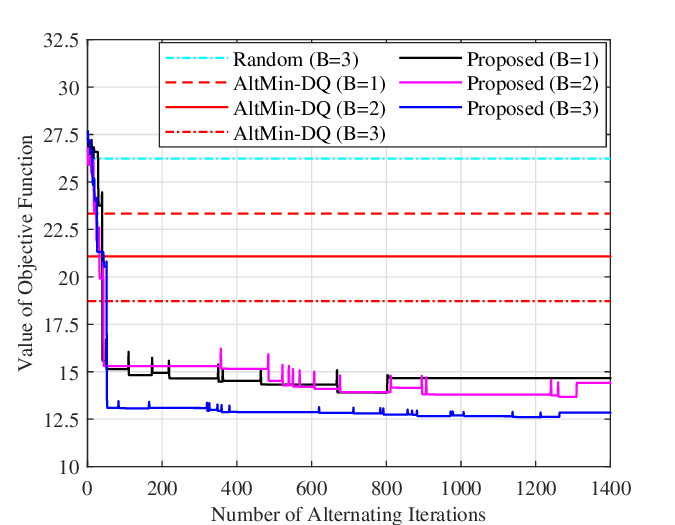}}
		\caption{Value of objective function against the number of alternating iterations between analog and digital sensing matrix at UE when ${N_{\text{r}}} = 32$, $N_\text{s}=N^\text{RF} = 4$, $T_\text{r} = 24$, ${G_{\text{r}}} = 36$ and $B = 1,2,3$.}
		\label{fig7}
	\end{figure}
	\begin{figure}[t]
		\center{\includegraphics[width=0.5\textwidth]{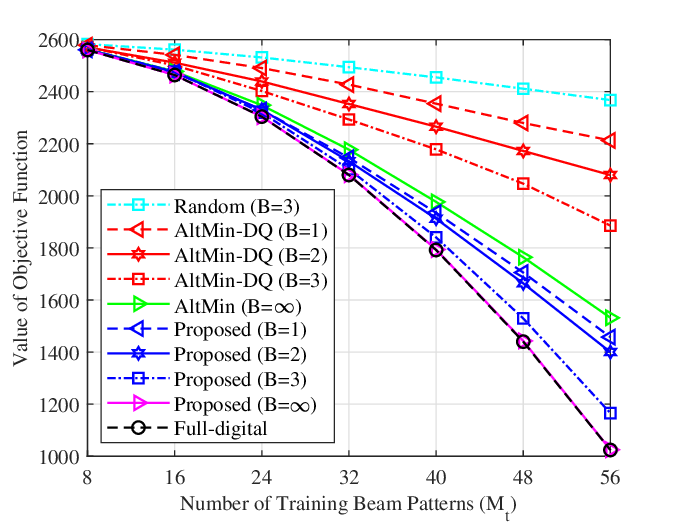}}
		\caption{Value of objective function against the number of training beam patterns at BS when ${N_{\text{t}}} = 64$, ${N_{\text{r}}} = 32$, $N_\text{s}=N^\text{RF} = 4$, $T_\text{t} = 2T_\text{r}$, ${G_{\text{t}}} = 72$, ${G_{\text{r}}} = 36$ and $B = 1,2,3,\infty$.}
		\label{fig8}
	\end{figure}

	In order to illustrate the convergence of the proposed Algorithm 3, the value of objective function (\ref{Blk_opt_form_a}) against the number of alternating iterations between ${\bf{W}}_{\text{RF},q}$ and ${\bf{W}}_{\text{BB},q}$ is shown in Fig. \ref{fig7}.
	There is no alternating iterations between analog and digital sensing matrix in the AltMin-DQ and random scheme, and we draw the straight lines with $B = 1,2,3$ as benchmarks for the performance curves of proposed Algorithm 3.
	Fig. \ref{fig7} shows that the proposed Algorithm 3 has fast and obvious convergence under the condition of low-resolution PSs on the whole, and it can achieve better performance than the AltMin-DQ scheme, where the performance curves slightly jitter due to the inevitable performance degradation caused by phase quantization during the optimization process of low-resolution analog precoder design in each block.
	Besides, in some specific interval after about 50 alternating iterations, the objective function within some consecutive blocks may not be reduced (since the number of consecutive blocks is less than $N_\text{r}^\text{Block}$, the stopping criterion does not triggered), so the sensing submatrices in the corresponding blocks are not updated, which can explain why the performance curve remains stable within some specific interval.

	In the previous optimization process, the sensing matrices are decoupled to be designed separately according to the symmetry  between BS and UE sides.
	To effectively show the coherence-based performance of the final equivalent dictionary matrix $\bf{Q}$ defined in (\ref{Q_def}) (the Kronecker product of the equivalent dictionary matrices at both sides), the equivalent objective function here can be defined by combining both BS and UE parts as
	\begin{align}\label{Q_obj_func}
		\left\| {\zeta {{\bf{Q}}^H}{\bf{Q}} - {{\bf{I}}_{{G_\text{t}}{G_\text{r}}}}} \right\|_F^2.
	\end{align}	
	Since the Gram matrix of $\bf{Q}$ is far from the identity matrix, we have to introduce a coefficient $\zeta$ to scale it, so that the objective function can reflect the effect of MIP optimization accurately.
	Note that this scaling will not change the coherence of any two columns of $\bf{Q}$.
	By expanding the Frobenius norm using the trace operator and calculating the gradient with respect to $\zeta$, the value of $\zeta$ can be given by
	\begin{align}\label{zeta_value}
		\zeta  = \frac{{{\rm{Tr}}\left\{ {{{\bf{Q}}^H}{\bf{Q}}} \right\}}}{{{\rm{Tr}}\left\{ {{{\bf{Q}}^H}{\bf{Q}}{{\bf{Q}}^H}{\bf{Q}}} \right\}}}.
	\end{align}	

	To verify the universality under different system parameters of the proposed algorithms, Fig. \ref{fig8} shows the value of objective function (\ref{Q_obj_func}) against the number of training beam patterns $T_\text{t}$.
	As shown in Fig. \ref{fig8}, as the number of training beam patterns increases, the corresponding overall coherence of $\bf{Q}$ decreases.
	
	\section{Conclusions}\label{S7}

	This paper considered the training beam design for channel estimation in hybrid analog-digital mmWave MIMO systems.
	We proposed two efficient iterative algorithms suitable for infinite-resolution and low-resolution PSs scenarios, respectively.
	The analog sensing matrices under both two scenarios are obtained according to the theory of manifold optimization to tackle the non-convex constant modulus constraints, while the optimization of digital sensing matrices under two scenarios is different.
	The digital sensing submatrices in different blocks can be jointly optimized with infinite-resolution PSs.
	However, in order to effectively cope with the performance loss caused by discrete phase, a block-wise optimization form is proposed under the low-resolution PSs scenario, and the corresponding digital sensing submatrices are iteratively optimized using a gradient descent method with adaptive steps.
	Simulation results show that the sparse channel estimated by training the beams using the proposed hybrid sensing matrix design with infinite-resolution PSs can be nearly as accurate as that achieved in full-digital systems.
	Even in low-resolution PSs scenarios, the proposed training beam design can significantly achieve better performance than that of existing schemes.

	\bibliographystyle{IEEEtran}
	\bibliography{IEEEabrv,Refference}

	\begin{IEEEbiography}[{\includegraphics[width=1in,height=1.05in,clip,keepaspectratio]{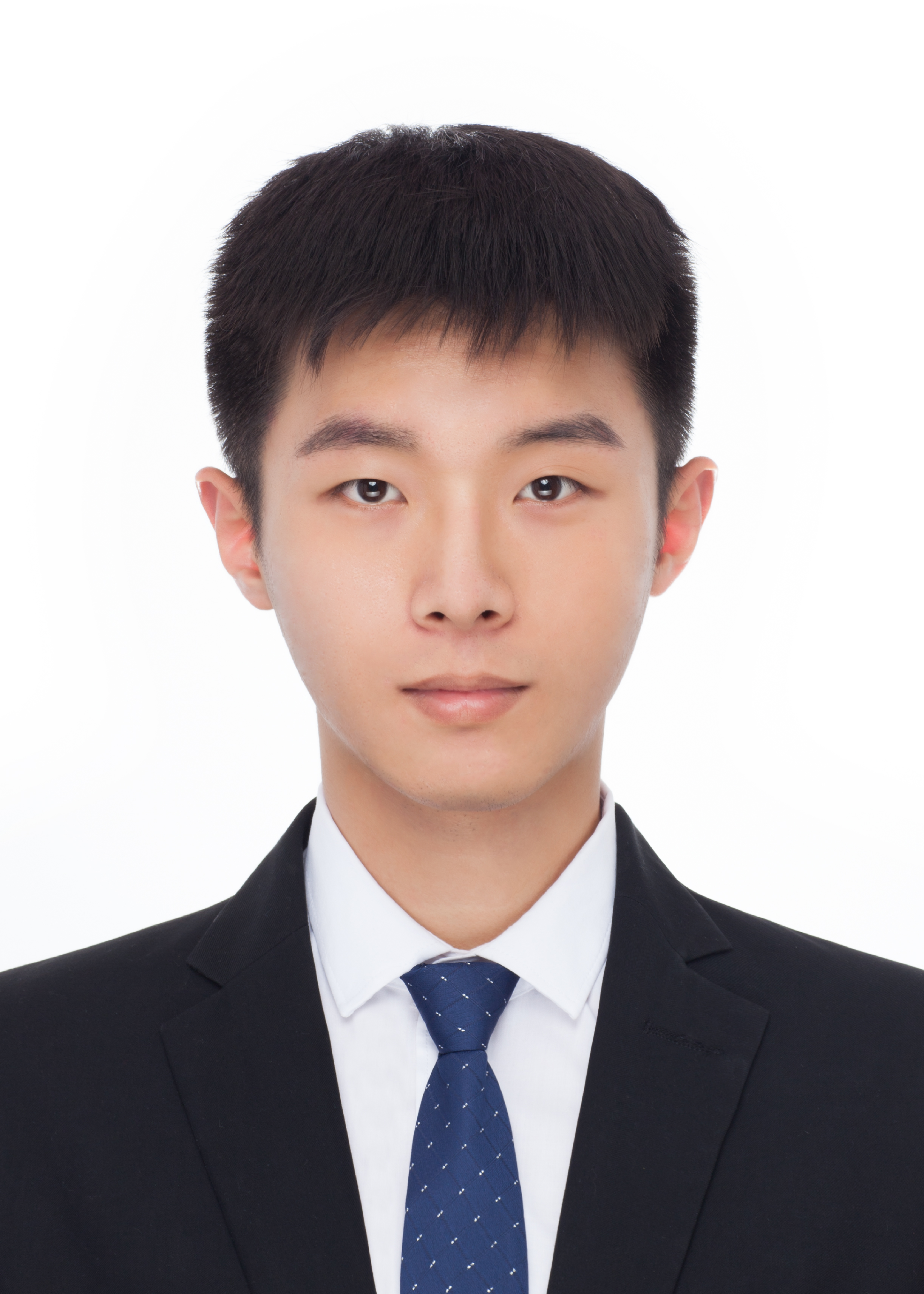}}]{Xiaochun Ge}
		received the B.S. degree in communication engineering from Beijing Institute of Technology, Beijing, China, in 2019, where he is currently pursuing the Ph.D. degree with the School of Information and Electronics. 	
		His main research interests include massive MIMO, intelligent reflecting surface, signal processing, and convex optimization.
	\end{IEEEbiography}

	\begin{IEEEbiography}[{\includegraphics[width=1in,height=1.05in,clip,keepaspectratio]{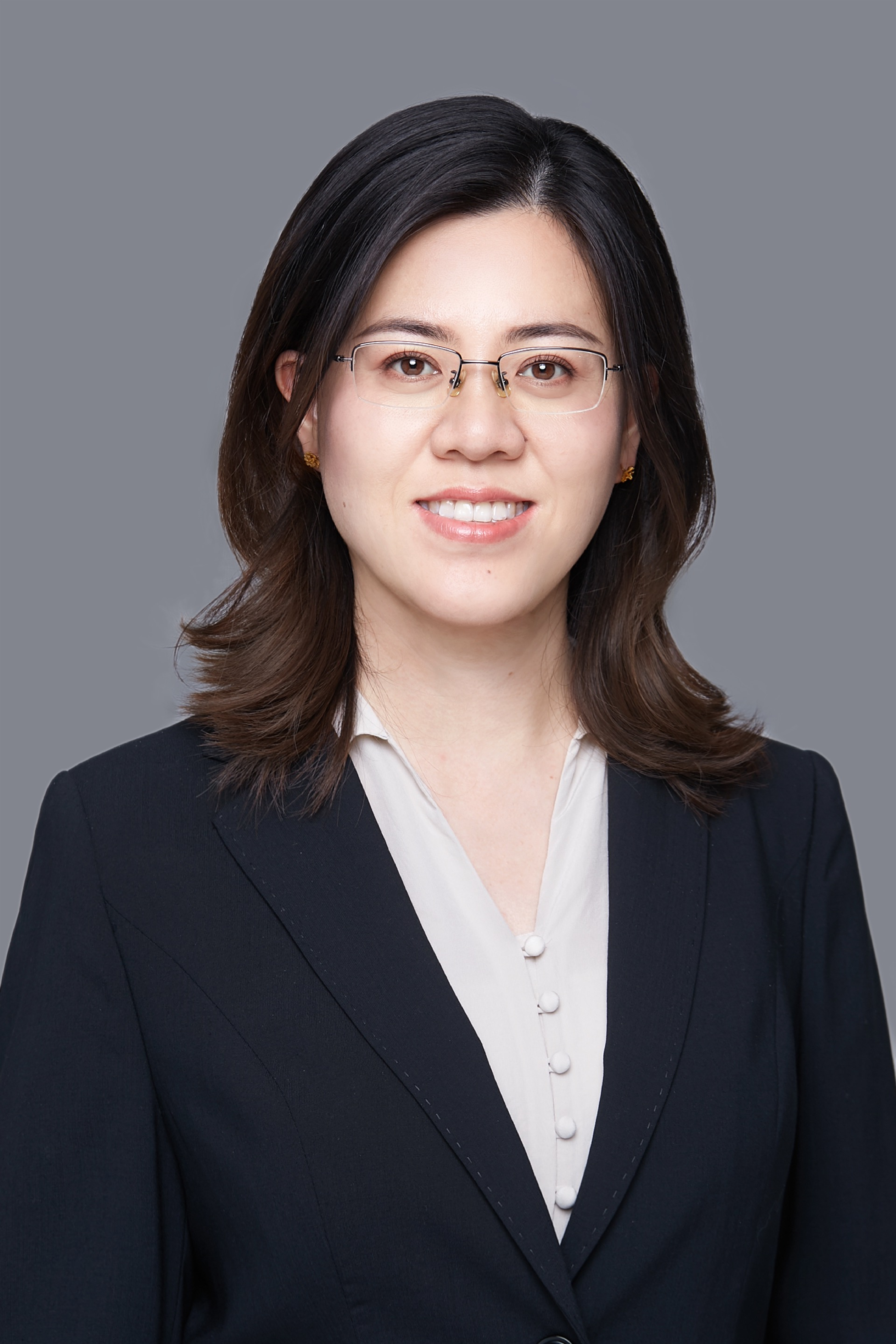}}]{Wenqian Shen}
		received the B.S. degree from Xi'an Jiaotong University, Shaanxi, China in 2013 and the Ph.D. degree from Tsinghua University, Beijing, China in 2018. She is currently an associate professor with the School of Information and Electronics, Beijing Institute of Technology, Beijing, China. Her research interests include massive MIMO and mmWave/THz communications. She has published several journal and conference papers in IEEE Transaction on Signal Processing, IEEE Transaction on Communications, IEEE Transaction on Vehicular Technology, IEEE ICC, etc. She has won the IEEE Best Paper Award at the IEEE ICC 2017.
	\end{IEEEbiography}

	\begin{IEEEbiography}[{\includegraphics[width=1in,height=1.05in,clip,keepaspectratio]{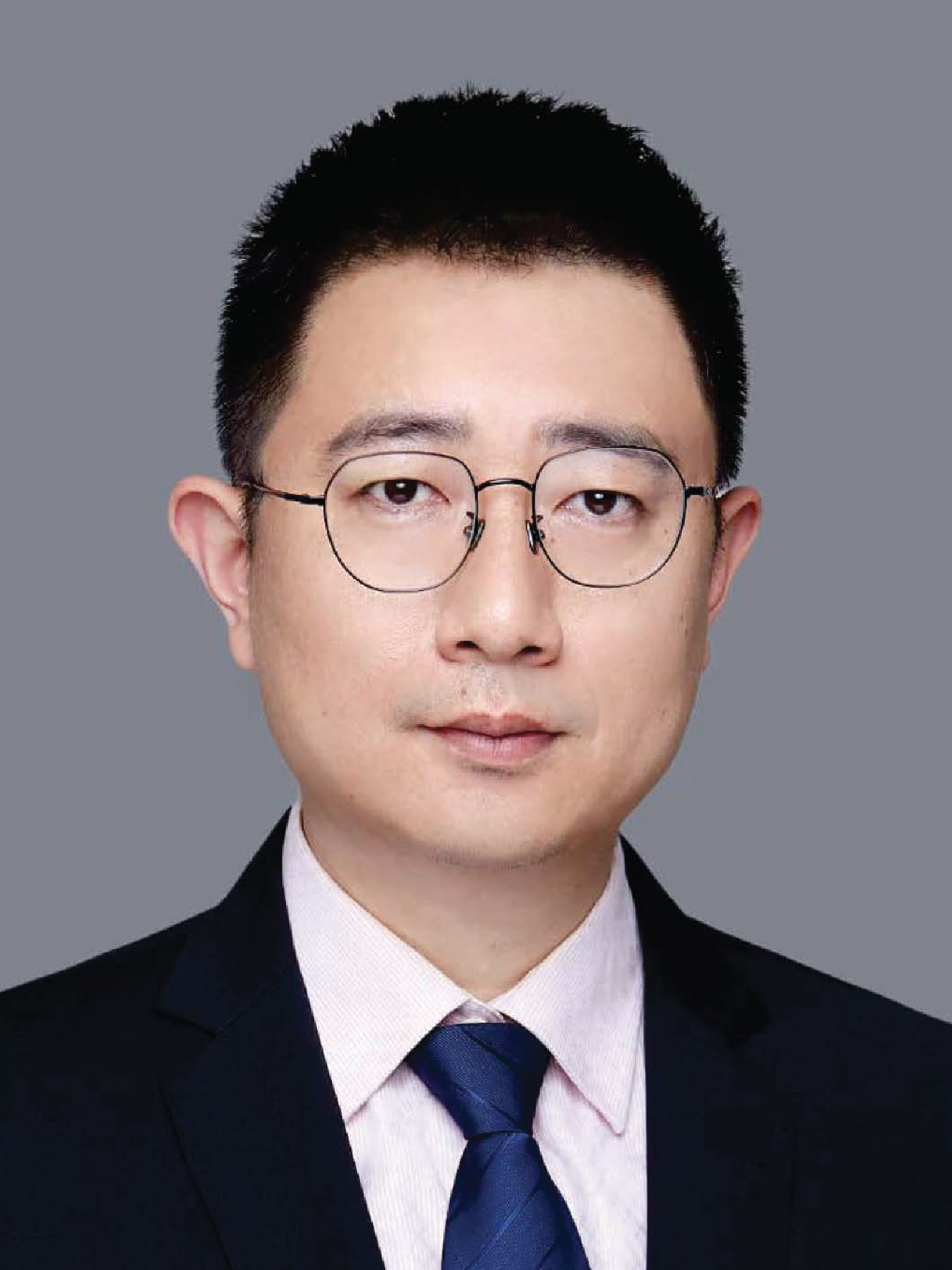}}]{Chengwen Xing}(Member, IEEE)
		received the B.E. degree from Xidian University, Xi’an, China, in 2005 and the Ph.D. degree from The University of Hong Kong, Hong Kong, China, in 2010. Since September 2010, he has been with the School of Information and Electronics, Beijing Institute of Technology, Beijing, China, where he is currently a Full Professor. From September 2012 to December 2012, he was a Visiting Scholar with the University of Macau, Macau, China. His current research interests include machine learning, statistical signal processing, convex optimization, multivariate statistics, array signal processing, and high-frequency band communication systems.
	\end{IEEEbiography}

	\begin{IEEEbiography}[{\includegraphics[width=1in,height=1.05in,clip,keepaspectratio]{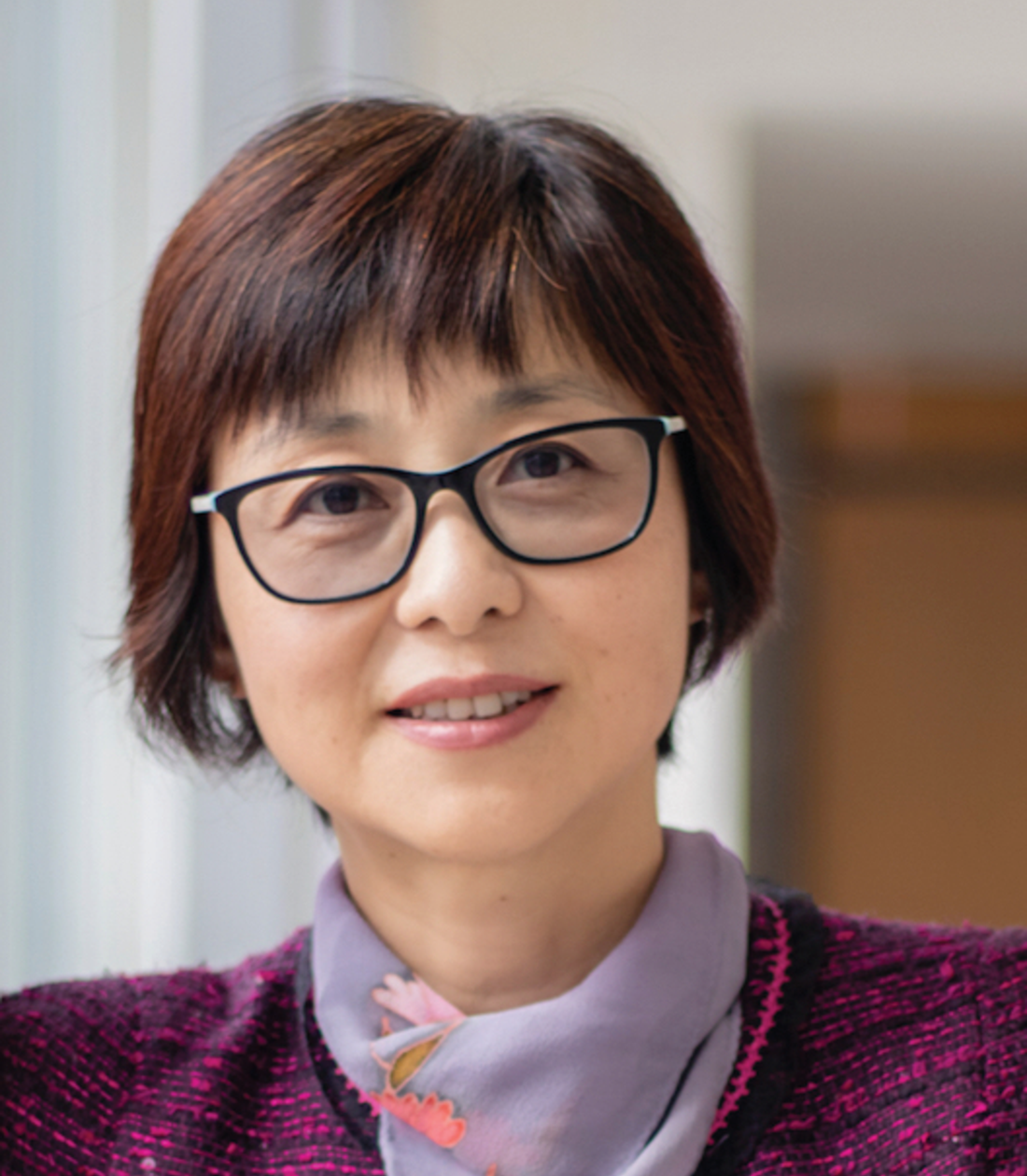}}]{Lian Zhao}(Senior Member, IEEE)
		received the Ph.D. degree from the Department of Electrical and Computer Engineering (ELCE), University of Waterloo, Canada, in 2002. She joined the Department of Electrical and Computer Engineering at Ryerson University, Toronto, Canada, in 2003 and a Professor in 2014. Her research interests are in the areas of wireless communications, resource management, mobile edge computing, caching and communications, and IoV networks. 

		She has been an IEEE Communication Society (ComSoc) Distinguished Lecturer (DL); received the Best Land Transportation Paper Award from IEEE Vehicular Technology Society in 2016, Top 15 Editor Award in 2016 for IEEE Transaction on Vehicular Technology, Best Paper Award from the 2013 International Conference on Wireless Communications and Signal Processing (WCSP), and the Canada Foundation for Innovation (CFI) New Opportunity Research Award in 2005. 
		She has been serving as an Editor for IEEE Transactions on Wireless Communications, IEEE Internet of Things Journal, and IEEE Transactions on Vehicular Technology (2013-2021). She served as a co-Chair of Wireless Communication Symposium for IEEE Globecom 2020 and IEEE ICC 2018; Finance co-Chair for 2021 ICASSP; Local Arrangement co-Chair for IEEE VTC Fall 2017 and IEEE Infocom 2014; co-Chair of Communication Theory Symposium for IEEE Globecom 2013. 
		
		She has severed as a panel expert for the Discovery Grant Program and Evaluation Committee for the Research Tools and Instruments Grants Program of Natural Sciences Engineering Research Council of Canada (NSERC). She is a licensed Professional Engineer in the Province of Ontario and a senior member of the IEEE Communication Society and Vehicular Technology Society. 
	\end{IEEEbiography}

	\begin{IEEEbiography}[{\includegraphics[width=1in,height=1.05in,clip,keepaspectratio]{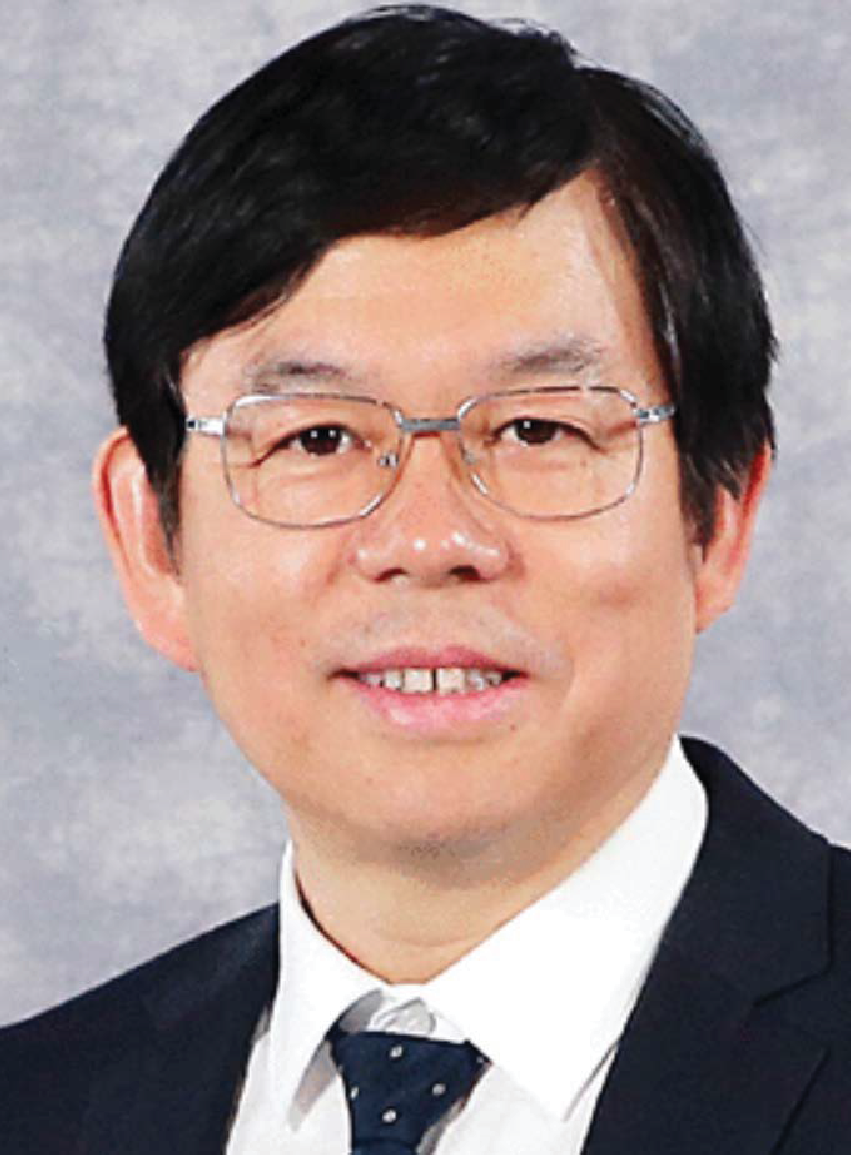}}]{Jianping An}(Member, IEEE)
		received the B.Eng. degree from PLA Information Engineering University, Zhengzhou, China, in 1987 and the Ph.D. degree from the Beijing Institute of Technology (BIT), Beijing, China, in 1996. From 2010 to 2011, he was a Visiting Professor with the University of California, San Diego, CA, USA. He is currently a Professor and the Dean of the School of Cyberspace Science and Technology in BIT. He has authored or coauthored more than 150 journal and conference articles, and holds or coholds 37 patents. His current research interests include signal processing theory and algorithms for communication systems. He was the recipient of a various awards for his academic achievements and the resultant industrial influences, including the National Award for Scientific and Technological Progress of China in 1997 and the Excellent Young Teacher Award by the China s Ministry of Education in 2000. Since 2010, he has been a Chief Reviewing Expert for the Information Technology Division, National Scientific Foundation of China.
	\end{IEEEbiography}

\end{document}